\documentclass[5p]{elsarticle}

\journal{Computational Materials Science}
\usepackage{amsbsy}
\usepackage{amsmath}
\usepackage{amssymb}
\usepackage{mathtools}
\usepackage{bm}
\usepackage[caption=false]{subfig}

\usepackage{booktabs}
\usepackage{tabularx}

\usepackage[utf8]{inputenc}
\usepackage{xcolor}

\providecommand{\unit}[1]{\ensuremath{\mathrm{\, #1}}}

\renewcommand{\vec}[1]{\boldsymbol{#1}}
\providecommand{\mat}[1]{\boldsymbol{#1}}
\providecommand{\nvec}[1]{\hat{\boldsymbol{#1}}}

\providecommand{\XX}{\mathbb{X}}
\providecommand{\YY}{\mathbb{Y}}
\providecommand{\ZZ}{\mathbb{Z}}

\begin{document}

\begin{frontmatter}

\title{Classification of atomic environments via the Gromov--Wasserstein distance}

\author{Sakura Kawano\corref{cor1}}
\address{Department of Chemical Engineering, University of California, Davis, Davis, CA 95616, USA}
\ead{skawano@ucdavis.edu}

\author{Jeremy K. Mason\corref{cor2}}
\address{Department of Materials Science and Engineering, University of California, Davis, Davis, CA 95616, USA}
\ead{jkmason@ucdavis.edu}

\cortext[cor2]{Corresponding author}

\begin{abstract}
Interpreting molecular dynamics simulations usually involves automated classification of local atomic environments to identify regions of interest. Existing approaches are generally limited to a small number of reference structures and only include limited information about the local chemical composition. This work proposes to use a variant of the Gromov--Wasserstein (GW) distance to quantify the difference between a local atomic environment and a set of arbitrary reference environments in a way that is sensitive to atomic displacements, missing atoms, and differences in chemical composition. This involves describing a local atomic environment as a finite metric measure space, which has the additional advantages of not requiring the local environment to be centered on an atom and of not making any assumptions about the material class. Numerical examples illustrate the efficacy and versatility of the algorithm.
\end{abstract}


\end{frontmatter}

% \linenumbers

\section{Introduction}
\label{sec:introduction}

Contemporary molecular dynamics simulations can involve millions of atoms, though the atoms participating in the phenomenon of interest (e.g., phase nucleation, shear band nucleation, surface adsorption) are generally many fewer. Some automated procedure to classify local atomic environments is therefore indispensable to initially identify these regions so that the researcher can perform additional analysis. Given the difficulty of precisely defining what an exceptional atomic environment would be in the absence of crystalline order, many of the procedures already proposed apply almost exclusively to crystalline solids. More specifically, the assumption is often made that most atoms are nearly on simple cubic (SC), body-centered cubic (BCC), face-centered cubic (FCC), or hexagonal close-packed (HCP) lattice sites, and the classification problem is reduced to assigning atoms to one of these classes (or to one other class containing all defected atomic environments).

Existing approaches can roughly be grouped as topological or geometric. Topological approaches construct either the network of bonds connecting neighboring atoms, or the Voronoi tessellation with the atomic positions as seeds. Atoms are assigned to a class by considering the number and arrangement of nearby bonds in the bond network or nearby faces of the Voronoi polyhedra. This intentionally disregards some information about the relative positions of the atoms to make the classification more robust to perturbations of the positions at finite temperatures (i.e., thermal noise). Topological approaches often have the advantages of computational efficiency, simplicity of exposition, and well-defined criteria for an atom to belong to a particular class. Examples in the literature include common neighbor analysis \cite{1987honeycutt,1994faken,2012stukowski}, crystal analysis \cite{2014stukowski}, neighborhood graph analysis \cite{2017reinhart}, Voronoi analysis \cite{1979hsu}, topological fingerprints \cite{2013schablitzki}, and Voronoi cell topology \cite{2015lazar}.

Geometric approaches instead map the relative positions of atoms in a local atomic environment to a continuous feature space. Each class is associated with a region of the feature space, and atoms whose feature vectors fall within one of these regions are assigned to that class. The regions are usually not defined a priori, but rather are constructed after observing the distribution of feature vectors of atoms in reference environments. Geometric approaches can provide information about the atomic environment that is not readily accessible to topological approaches, e.g., point symmetry groups or elastic strain tensors, but can suffer more from thermal noise and be more expensive to calculate. Examples in the literature include the centrosymmetry parameter \cite{1998kelchner}, bond-orientational order parameters \cite{1983steinhardt,2016winczewski}, the Minkowski structure metric \cite{2013mickel}, bond angle analysis \cite{2006ackland}, neighbor distance analysis \cite{2012stukowski}, and polyhedral template matching \cite{2016larsen}.

Of the approaches above, adaptive common neighbor analysis (ACNA) \cite{2012stukowski} and polyhedral template matching (PTM) \cite{2016larsen} are perhaps the most frequently used to identify atomic environments in crystalline solids. They perform particularly well for molecular dynamics simulations of single-component systems, and the procedure proposed here is not necessarily intended for such applications. That said, there are still several respects in which they could be improved.

First, they are effectively limited to consider only one or two nearest neighbor shells around a central atom. This is a consequence of the way the local bond network is constructed for ACNA, and of the use of a convex hull as part of the matching algorithm for PTM. As the accuracy of interatomic potentials in two and three component systems continues to improve and simulations of materials with more complex crystal structures become more common, methods able to handle extended environments will likely become more relevant.

Second, the methods are sensitive to atoms entering or leaving the local environment; this is related but not entirely equivalent to being robust to thermal noise. ACNA reduces the frequency of such events by varying the radius of the local environment with the reference environment and the atomic positions, while PTM uses a topological ordering of nearby atoms to make the classification resistant to perturbations in the atomic positions. Nevertheless, a shear strain applied to a large atomic environment could still displace some of the atoms enough to leave the region being considered and frustrate the analysis.

Third, they can only include limited information about the chemical composition of the local environment, at least in the forms currently in the literature. ACNA could be adapted to include chemical information by appending the species of the atoms along bond chains \cite{2007lummen}, though this would be unwieldy for three or more chemical species. PTM has been used for binary alloys \cite{2016larsen}, but apparently requires considerable symmetry in the arrangement of the chemical species. A more flexible approach would be valuable, particularly if molecular dynamics simulations of two- and three-component systems become more common.

The procedure proposed here is based on the Gromov--Wasserstein (GW) distance recently defined by Memoli \cite{2007memoli,2011memoli,2017memoli}, which up to now has mostly been used for shape matching in the field of computer vision \cite{2013schmitzer,2016solomon,2016peyre}. For example, the GW distance can be used to match an object represented as an incomplete point cloud to one of a set of reference objects, perhaps in a difference pose. This is not dissimilar to matching a local atomic environment to one of a set of reference environments, possibly with perturbed atomic positions or some of the atoms missing. Apart from resolving the three limitations above, our approach has the additional advantages of not requiring the local environment to be centered on an atom (e.g., for the identification of vacancies) and of providing a metric on the space of all local atomic environments. That said, the GW distance is more complicated to define and is substantially slower to calculate than ACNA and PTM, and for that reason is intended to be complementary to them.

\section{Finite Metric Measure Spaces}
\label{sec:finite_metric_measure_spaces}

\begin{figure}
\includegraphics[width=\columnwidth]{%
	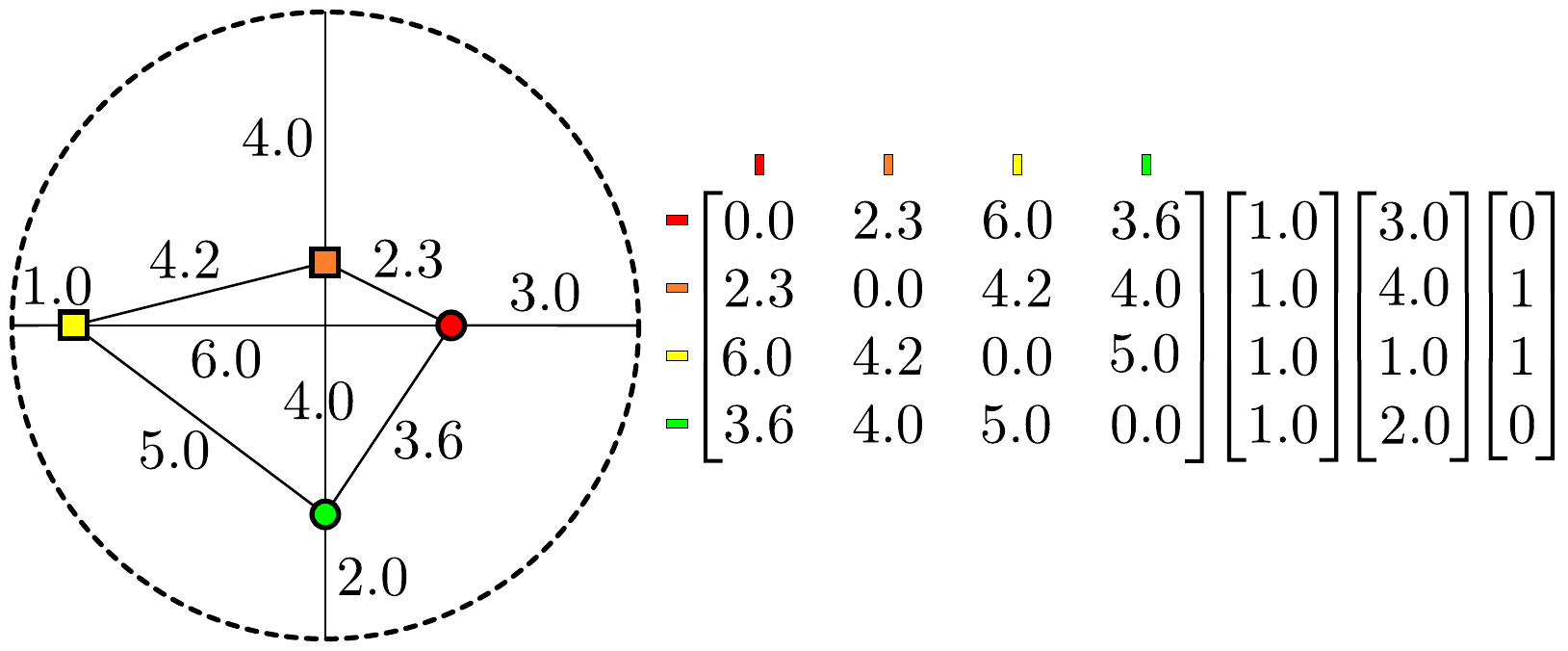}
\caption{\label{fig:finite_space}A description of a local atomic environment as a finite metric measure space. Color indicates the distinct points of the finite space, and circles and squares indicate the two chemical species. Numbers in the local atomic environment (left) are dimensionless Euclidean distances. The metric (second from left) indicates the pairwise distances between atoms, the measure (middle) indicates the fraction of an atom associated with each point, the distances to the boundary (second from right) are the distances from each atom to the closest point on the boundary, and the species labels (right) indicate the chemical species of the atoms.}
\end{figure}

A local atomic environment is often described by a set of vectors from the central atom to the surrounding atoms. The GW distance instead requires that a local atomic environment be described as a \emph{finite metric measure space}. As the name implies, this involves the construction of a finite space, a metric describing distances in the space, and a measure describing the distribution of atoms in the space. Figure \ref{fig:finite_space} is a concrete example of the construction for a spherical region. While the region is not required to be spherical, this simplifies some of the analysis and will be assumed throughout.

A \emph{finite space} is a topological space that contains only a finite number of points. The natural choice for a local atomic environment is one point for each atomic center, as indicated by the red, orange, yellow and green points in Figure \ref{fig:finite_space}. For our purposes, a \emph{metric} is a symmetric matrix of pairwise Euclidean distances between points, and a \emph{measure} is a function that assigns values to points. The GW distance uses these as weights to indicate the relative importance of points in the space, but otherwise does not specify their interpretation. Here, the measure will be used to indicate the number of atoms associated with a point. Since atoms are indivisible and the position of each atom is unique, all of the entries will be $1.0$.

While not part of the definition of a finite metric measure space, our description of a local atomic environment includes a vector of \emph{distances to the boundary} for each atom and a vector of \emph{species labels} that indicates the chemical species of the atoms associated with each point. Distances to the boundary are used to penalize the departure of atoms from the environment. This is envisioned as involving the motion of an atom to the boundary, and hence is proportional to the distance to the boundary. The chemical species is well-defined since each point is associated with a single atom. By convention, the chemical species are labeled with increasing integers starting with zero. 

Describing a local atomic environment as a finite metric measure space instead of as a set of bond vectors has several advantages. First, Figure \ref{fig:finite_space} shows that the local atomic environment does not need to be centered on an atom. This allows the GW distance to be used to find, e.g., the precise locations of vacancies or interstitial sites in a finite temperature system. Second, the distance matrix is invariant to translations, rotations, and reflections of the local atomic environment; these symmetries do not need to be handled in a separate calculation as with PTM.

\section{Gromov--Wasserstein Distance}
\label{sec:gromov_wasserstein}

The GW distance is a metric \cite{2011memoli} that allows the comparison of finite metric measure spaces. More specifically, let $X$ be a finite space with metric $\mat{d}^X$ and measure $\vec{\mu}^X$; the triple $\XX = \{X, \mat{d}^X, \vec{\mu}^X\}$ is a finite metric measure space. The GW distance is then a function $\mathcal{G}(\XX, \YY)$ with the following properties for all finite metric measure spaces $\XX$, $\YY$ and $\ZZ$ with the same total measures:
\begin{enumerate}
\item $\mathcal{G}(\XX, \YY) \geq 0$,
\item $\mathcal{G}(\XX, \YY) = 0$ if and only if $\XX = \YY$,
\item $\mathcal{G}(\XX, \YY) = \mathcal{G}(\YY, \XX)$, and
\item $\mathcal{G}(\XX, \ZZ) \geq \mathcal{G}(\XX, \YY) + \mathcal{G}(\YY, \ZZ)$.
\end{enumerate}
These conditions are designed to ensure that every metric conform to our usual intuitions about distance in Euclidean space. In particular, the fourth condition is known as the triangle inequality, and is required for the clustering of points to be defined in a meaningful way; without this, even if $\XX$ is close to $\YY$ and $\YY$ is close to $\ZZ$, $\XX$ and $\ZZ$ could still be arbitrarily far apart. The definition of a metric is provided here because our use of the word is somewhat more restricted than elsewhere in the materials science literature \cite{2011keys}.

The notion of a \emph{measure coupling} will be useful when describing the calculation of the GW distance. Given finite metric measure spaces $\XX$ and $\YY$ with $n$ and $m$ points, an admissible measure coupling between them is an $n \times m$ matrix $\mat{\mu}$ with non-negative entries. Intuitively, this provides a correspondence of points in $X$ with points in $Y$ that allows for partial matching. Denote the row and column sums as $\nu^{X}_{i} = \sum_{j} \mu_{ij}$ and $\nu^{Y}_{j} = \sum_{i}\mu_{ij}$ for all $\textit{i} \in$ $[1, \textit{n}]$ and $\textit{j} \in$ $[1, \textit{m}]$. A measure coupling can be balanced or unbalanced, where a balanced measure coupling is one for which the row sums equal $\vec{\mu}^X$ and the column sums equal $\vec{\mu}^Y$, i.e.,  $\nu^{X}_{i} = \mu^{X}_{i}$ and $\nu^{Y}_{j} = \mu^{Y}_{j}$. Let the set of all admissible unbalanced measure couplings for the finite metric measure spaces $\XX$ and $\YY$ be indicated by $\mathcal{M}(\vec{\mu}^X, \vec{\mu}^Y)$. 

Given an admissible measure coupling $\mat{\mu}$, define the quantity
\begin{equation*}
J(\mat{\mu} | \mat{d}^X, \mat{d}^Y) = \sum_{i', i = 1}^{n} \sum_{j', j = 1}^{m} |d^X_{i'i} - d^Y_{j'j}| \mu_{i'j'} \mu_{ij}
\end{equation*}
and let $\lambda^{X}_{i}$ be the distance to the boundary of the $i$th point of $X$. Then the unbalanced GW distance\footnote{This is actually the unbalanced $1$-Gromov--Wasserstein distance. $p$-Gromov--Wasserstein distances can be defined for any $p \in [1, \infty)$.} between $\XX$ and $\YY$ is defined here as

\begin{align}
\mathcal{G}(\XX, \YY) = \min_{\mat{\mu} \in \mathcal{M}} &\bigg[\frac{1}{2} J(\mat{\mu} | \mat{d}^X, \mat{d}^Y) + \sum_{i = 1}^{n} \lambda^{X}_{i} \big| {\nu^{X}_{i} -\mu^{X}_{i}} \big|  \nonumber\\
& + \sum_{j = 1}^{m} \lambda^{Y}_{j} \big| {\nu^{Y}_{j} -\mu^{Y}_{j}} \big| \bigg],
\label{eq:gromov_wasserstein}
\end{align}
following the same approach as for the unbalanced Wasserstein distance of Chizat et al. \cite{2018chizat}. The motivation for the unbalanced GW distance is that it is not always possible to find a balanced measure coupling, e.g., when there are unequal number of atoms between the reference and local environments. 

\begin{figure*}
\centering
	\includegraphics[width=0.84\textwidth]{%
	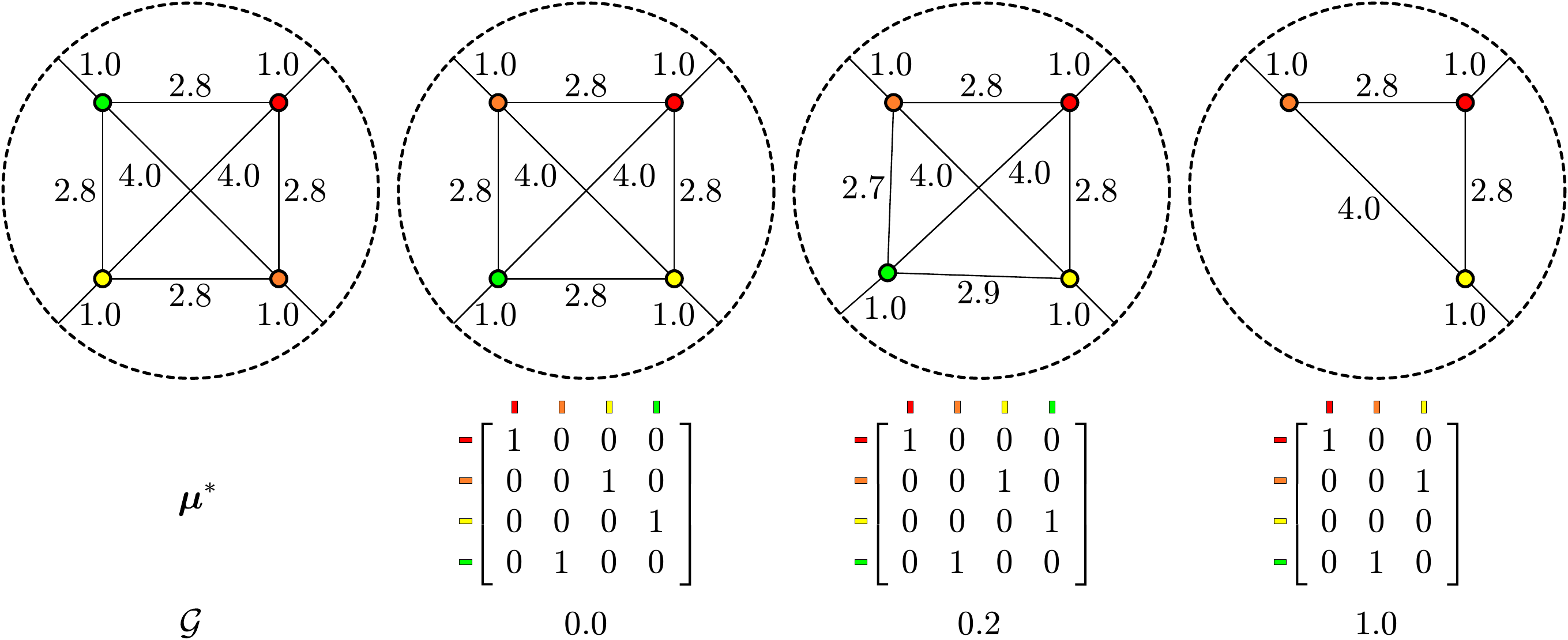}
\caption{\label{fig:intuition}Examples of local atomic environments intended to clarify the meaning of Eq.\ \ref{eq:gromov_wasserstein}. The leftmost environment $\XX$ is the reference environment, with the numbers indicating the distances between pairs of atoms. The second from the left environment $\YY_1$ differs only by a permutation of the atomic labels, the second from the right environment $\YY_2$ additionally has a perturbation applied to the green atom's position, and the rightmost environment $\YY_3$ instead has a missing atom. The second row gives the measure couplings $\mat{\mu}^*$ that realize the minimum of Eq.\ \ref{eq:gromov_wasserstein}, with the rows corresponding to atoms of $\XX$ and the columns to atoms of $\YY_i$. The third row gives the GW distance $\mathcal{G}(\XX, \YY_i)$ to the same precision as the distances.}
\end{figure*}

Figure \ref{fig:intuition} provides several examples that are intended to help the reader develop an intuition for this definition. Let the reference environment $\XX$ be the leftmost in the figure. The local atomic environment $\YY_1$ second from the left is identical to $\XX$ except for a permutation of the atomic labels, and the $\mat{\mu}^*$ that achieves the minimum in Eq.\ \ref{eq:gromov_wasserstein} is a permutation matrix that maps one set of atoms to the other (e.g., the orange atom in $\XX$ is mapped to the yellow atom in $\YY_1$). Since $\XX$ and $\YY_1$ differ only by a symmetry of the physical system, the GW distance $\mathcal{G}(\XX, \YY_1)$ vanishes. The local atomic environment $\YY_2$ second from the right additionally has a small perturbation applied to the green atom's position, visible as changes of distance to the yellow and orange atoms. In this case $\mat{\mu}^*$ remains the same, but $\mathcal{G}(\XX, \YY_2)$ is the sum of the magnitudes of the distance changes between every pair of atoms. Since the distances in $\XX$ and $\YY_2$ are the same except for those around the perturbed atom, $\mathcal{G}(\XX, \YY_2)$ is the sum of the magnitudes of the distance changes from the green to the yellow and orange atoms in $\YY_2$. The rightmost local atomic environment $\YY_3$ instead has a missing atom, requiring that the corresponding column of $\mat{\mu}^*$ be removed. The resulting discrepancy between $\mat{\mu}^X$ and $\mat{\nu}^X$ makes $\mathcal{G}(\XX, \YY_3)$ the distance the missing atom would have traveled to reach the boundary and leave the local atomic environment. The minimum in Eq.\ \ref{eq:gromov_wasserstein} allows the GW distance to remain continuous as the magnitude of the perturbation increases and removing the atom becomes the less expensive option.

There are several other conditions that should be satisfied by the finite metric measure spaces before the GW distance is applied. First, the measures should be strictly positive. Any points for which the measures are zero (i.e., that are not occupied by atoms) should be removed from the spaces, the corresponding rows and columns removed from the distance matrices, and the corresponding entries removed from the measures, distances to the boundary, and species labels. Second, the algorithm is more stable when the median off-diagonal entry of the distance matrices is of order one. If $\kappa$ is the median of these entries, then the distance matrices and distances to the boundary should be divided by $\kappa$ before the calculation, and the GW distance multiplied by $\kappa$ after the calculation.

This leaves the problem of finding a measure coupling $\mat{\mu}^*$ that realizes the minimum in Eq.\ \ref{eq:gromov_wasserstein}. Formally, this is at least as difficult as a nonconvex quadratic optimization problem with linear constraints, and for which there is no known polynomial-time algorithm to find the global minimum \cite{1991pardalos}. In practice, the approach followed in the literature \cite{2007memoli,2011memoli,2016solomon,2016peyre} is to approximate $\mat{\mu}^*$ by successive linear optimization problems, and the same approach is followed here:
\begin{enumerate}
\item Initialize $\mat{\mu}$ with some admissible measure coupling $\mat{\mu}^0$ and set $k = 0$.
\item Solve the linear optimization problem
\begin{align}
c^{k}_{ij} &= \frac{1}{2} \sum_{i' = 1}^{n} \sum_{j' = 1}^{m} |d^X_{ii'} - d^Y_{jj'}| \mu^{k}_{i'j'}, \nonumber \\
\mat{\mu}^{k + 1} &= \underset{\mat{\mu} \in \mathcal{M}} {\mathrm{argmin}} \bigg[ \sum_{i = 1}^{n} \sum_{j = 1}^{m} c^{k}_{ij} \mu_{ij} + \sum_{i = 1}^{n} \lambda^{X}_{i} \big| \nu^{X}_{i} - \mu^{X}_{i} \big| \nonumber \\
&\hspace{45pt} + \sum_{j = 1}^{m} \lambda^{Y}_{j} \big| \nu^{Y}_{j} - \mu^{Y}_{j} \big| \bigg].
\label{eq:wasserstein}
\end{align}
\item If the stopping criterion is satisfied, set $\mat{\mu}^* = \mat{\mu}^{k + 1}$ and exit. If not, set $k = k + 1$ and return to Step 2.
\end{enumerate}
This is known as the alternate convex search algorithm \cite{2011memoli,2007gorski}, and converges to a local minimum of the original problem. In principle, the quality of the result could be improved by repeatedly running the algorithm with randomized initial conditions. In practice, the measure coupling is  initialized to a constant matrix and heuristic perturbations are regularly applied to break any symmetries. The efficacy of this approach is visible in Section \ref{sec:MD}.

The linear optimization problem in Eq.\ \ref{eq:wasserstein} is identical to the one used to calculate an unbalanced Wasserstein distance \cite{2018chizat}. Let $\epsilon > 0$ be a regularization parameter and replace the linear optimization problem in Eq.\ \ref{eq:wasserstein} with
\begin{align}
\mat{\tilde{\mu}}^{k + 1} = \underset{\mat{\mu} \in \mathcal{M}}							  {\mathrm{argmin}} &\bigg[ \sum_{i = 1}^{n} \sum_{j = 1}^{m} (c^{k}_{ij} + \epsilon \log \mu_{ij}) \mu_{ij}   \nonumber\\
& +  \sum_{i = 1}^{n} \lambda^{X}_{i} \big| \nu^{X}_{i} - \mu^{X}_{i} \big| +  \sum_{j = 1}^{m} \lambda^{Y}_{j} \big| \nu^{Y}_{j}- \mu^{Y}_{j} \big| \bigg] \nonumber.
\end{align}
This can be solved efficiently with a modified Sinkhorn-Knopp algorithm \cite{2018chizat} as follows:
\begin{enumerate}
\item Initialize $a^0_i = 1$ for all $i \in [1, n]$, $b^0_j = 1$ for all $j \in [1, m]$, $\gamma_{ij} = \exp(- c^{k}_{ij}/\epsilon)$, and set $\ell = 0$.
\item Set $a^{\ell + 1}_i = \min[e^{\lambda^{X}_i / \epsilon}, \max(e^{-\lambda^{X}_i / \epsilon}, \mu^X_i / \sum_j \gamma_{ij} b^{\ell}_j ) ]$.
\item Set $b^{\ell + 1}_j = \min[e^{\lambda^{Y}_j / \epsilon}, \max(e^{-\lambda^{Y}_j / \epsilon}, \mu^Y_j / \sum_i a^{\ell + 1}_i \gamma_{ij} )]$
\item If the stopping criterion is satisfied, set $\mu_{ij}^{k + 1} = \\
a_i^{\ell + 1} \gamma_{ij} b_j^{\ell + 1}$ and exit. If not, set $\ell = \ell + 1$ and return to Step 2.
\end{enumerate}
Decreasing $\epsilon$ reduces the regularization and drives $\mat{\tilde{\mu}}^{k + 1}$ toward the solution of Eq.\ \ref{eq:wasserstein}, but can introduce numerical instabilities. 

Our implementation uses the log-domain stabilization and $\epsilon$-scaling of Schmitzer \cite{2016schmitzer,2018chizat}. These modifications to the basic Sinkhorn-Knopp algorithm require the introduction of several additional parameters; $\tau = 1000$ regulates the frequency of absorption iterations for the log-domain stabilization, and $\epsilon$ is scaled by factors of $4$ from an initial value of the median of the $c_{ij}$ to a final value of $0.001$ times the median distance between distinct points. The various heuristics used to escape local minima are described in \ref{sec:heuristics}. The algorithm is written as a library in portable C11 with Python and MATLAB interfaces, is open source, and is available on request.

\section{Classification of Atomic Environments}
\label{sec:classification}

While the calculation of the unbalanced GW distance introduced in Section \ref{sec:gromov_wasserstein} is relevant to general finite metric measure spaces, this section instead describes the application of the unbalanced GW distance to the classification of local atomic environments. In particular, Section \ref{sec:gromov_wasserstein} does not introduce the chemical species of the atoms. The extension of the unbalanced GW distance to the case of multiple species is called the composition-restricted Gromov--Wasserstein (CRGW) distance.

First, the user should specify a region of Euclidean space to be used for the definition of all local atomic environments. Since crystal structure and orientation can vary throughout a simulation cell, a spherical region with a radius of $1.5$ to $2.5$ times the average atomic spacing is a reasonable choice. 

Second, the user should provide a reference atomic environment for each class being considered. The finite metric measure spaces of the reference atomic environments are then constructed and stored for subsequent use. These take the form of sets $\mathbb{X} = \{X, \mat{d}^X, \vec{\mu}^{X}, \vec{\lambda}^{X}, \vec{\delta}^{X}\}$ where $\vec{\lambda}^{X}$ and $\vec{\delta}^{X}$ contain the distances to the boundary and the species labels.

Third, the local atomic environment to be classified is identified, and the corresponding finite metric measure space $\mathbb{Y} = \{Y, \mat{d}^Y, \vec{\mu}^{Y}, \vec{\lambda}^{Y}, \vec{\delta}^{Y}\}$ is constructed using the same region as before. The CRGW distance from $\YY$ to each of the reference environments is calculated, and a user-specified criterion is used to classify the local atomic environment on the basis of these distances. Part of the advantage of this approach is that the classification criterion can be as simple or as complex as the user desires; the local atomic environment could be assigned to the class with the smallest distance, or assigned to the most likely class using the probability distributions of distances developed in Section \ref{sec:thermal_noise}.

This leaves the calculation of the CRGW distance itself. Let $\XX$ and $\YY$ be the finite metric measure spaces of the reference and local atomic environments, and have $n$ and $m$ atoms respectively. Let $\mat{R}$ be an $n \times m$ matrix with $R_{ij}$ equal to one if the $i$th atom of $\XX$ and the $j$th atom of $\YY$ have the same species label, and zero otherwise. Then the CRGW distance $\mathcal{D}(\XX, \YY)$ is still defined by means of Eq.\ \ref{eq:gromov_wasserstein}, but with the minimization performed over the restricted set of measure couplings with the same zero entries as $\mat{R}$ (atoms of different chemical species cannot be coupled). Within the context of Section \ref{sec:gromov_wasserstein}, this restriction can be realized by replacing the initial measure coupling $\mu^0_{ij}$ with $R_{ij} \mu^0_{ij}$  and replacing $\gamma_{ij}$ in Step 1 of the modified Sinkhorn-Knopp algorithm with $\gamma_{ij} = R_{ij} \exp(-c_{ij}/\epsilon)$.

Note that the calculation of the CRGW distance actually increases in efficiency with the number of chemical species for a fixed number of atoms. The reason for this is that the sparsity of $\mat{R}$ increases with the number of chemical species, dramatically reducing the set of possible measure couplings in Eq.\ \ref{eq:gromov_wasserstein}. That said, any efficiency gains would likely be offset by an increase in the number of reference atomic environments defined by the user.

With the CRGW distance defined, the rest of this section consists of illustrative examples where the procedure is applied to local atomic environments in two dimensions. This simplification is used only for clarity of the figures; since the CRGW distance does not explicitly depend on the dimension of the ambient space, the calculation is precisely the same in two and three dimensions.

\begin{figure}
\includegraphics[width=\columnwidth]{%
	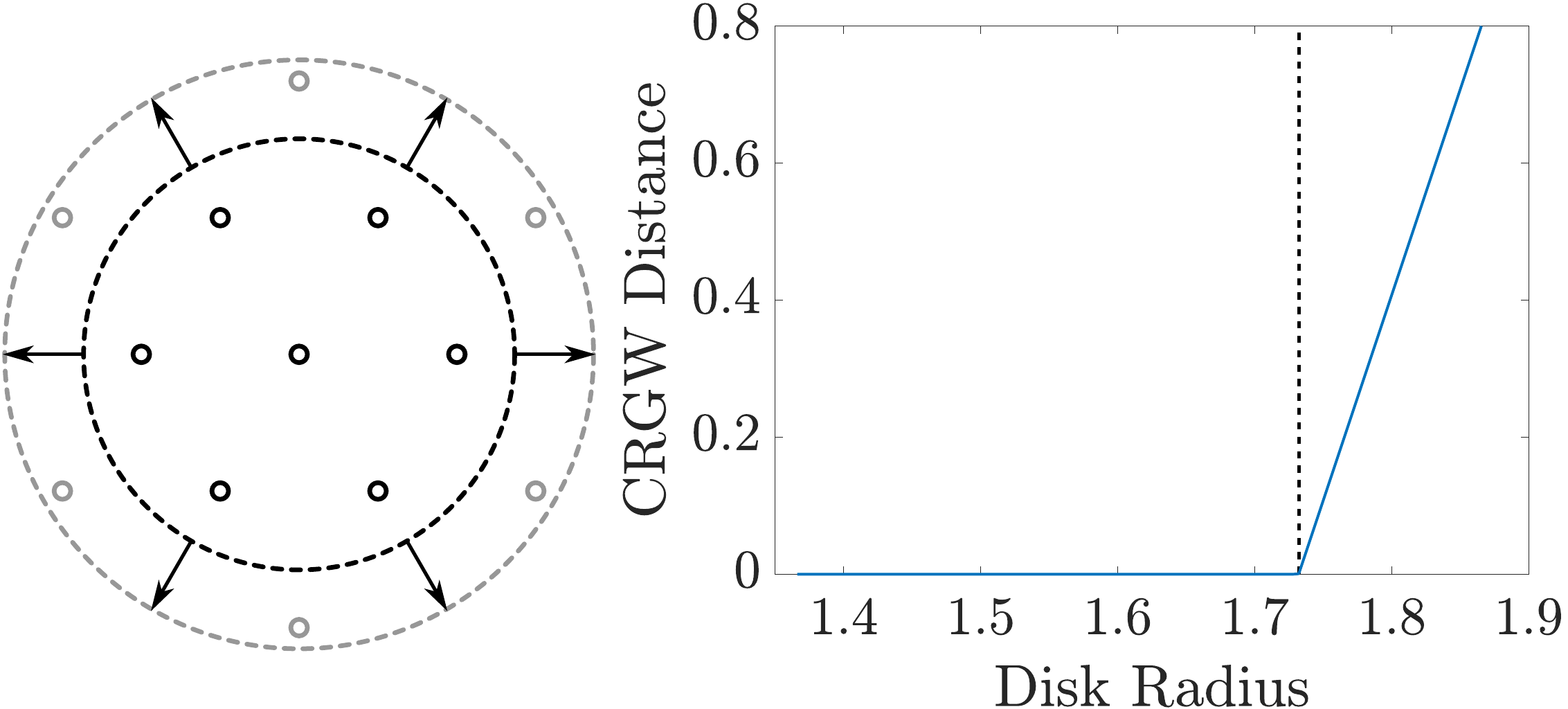}
\caption{\label{fig:change_radius}The CRGW distance is continuous with respect to atoms entering and leaving a local atomic environment. The radius of the environment on the left is increased from $(1 + \sqrt{3}) / 2$ to $(2 + \sqrt{3}) / 2$ in units of the atomic spacing. The distance to the initial condition on the right is continuous, with a discontinuous first derivative at $\sqrt{3}$.}
\end{figure}

Figure \ref{fig:change_radius} shows that the CRGW distance is continuous with respect to atoms entering and leaving the local atomic environment. The atoms are arranged on a triangular lattice with unit spacing, and the radius of the local atomic environment on the left is increased from $(1 + \sqrt{3}) / 2$ to $(2 + \sqrt{3}) / 2$. The CRGW distance to the environment is a continuous function of the radius, even though six atoms enter the region at a radius of $\sqrt{3}$ and cause a discontinuous derivative at the dashed vertical line. 

\begin{figure}
\includegraphics[width=\columnwidth]{%
	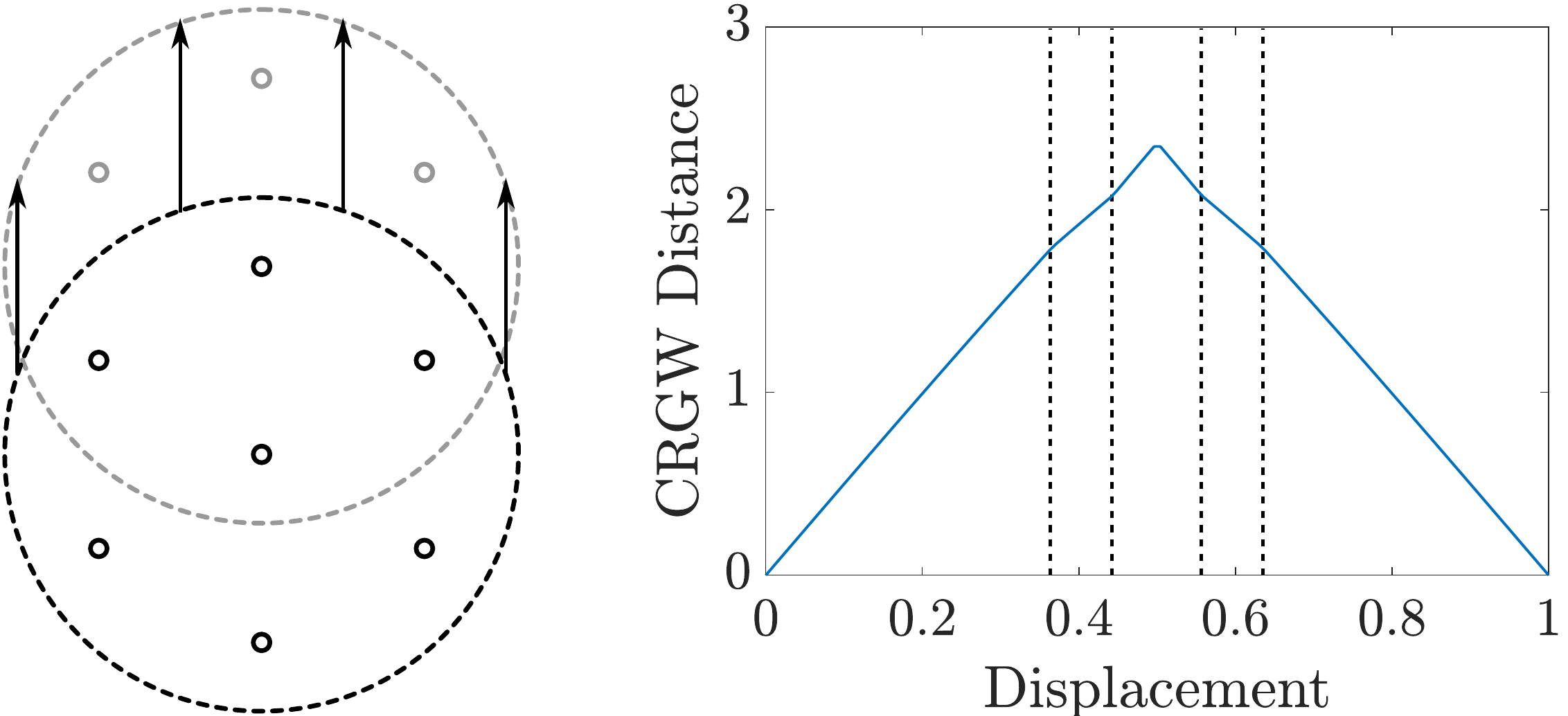}
\caption{\label{fig:change_center}The CRGW distance is continuous with respect to displacements of the local atomic environment. The center of the environment on the left is moved in the vertical direction by one atomic spacing. The distance to the initial condition on the right is continuous, with discontinuous first derivatives at $(\sqrt{3} - 1) / 2$ and $(3 - \sqrt{1 + 2 \sqrt{3}}) / 2$, and the reflection of these quantities about $0.5$.}
\end{figure}

Figure \ref{fig:change_center} shows that the CRGW distance is continuous with respect to displacements of the local atomic environment. The atoms are arranged on a triangular lattice with unit spacing as before, and the center of a local atomic environment of radius $(1 + \sqrt{3}) / 2$ is moved along a straight line between neighboring atoms. The CRGW distance to the initial environment is a continuous function of the displacement, passing through a maximum halfway between the atoms before returning to zero. The first two dashed vertical lines indicate discontinuous derivatives caused by the bottom atom leaving the environment and two of the uppermost atoms entering the environment, respectively. The environment briefly contains eight atoms before two corresponding events occur in reverse order as the distance returns to zero.

\begin{figure}
\includegraphics[width=\columnwidth]{%
	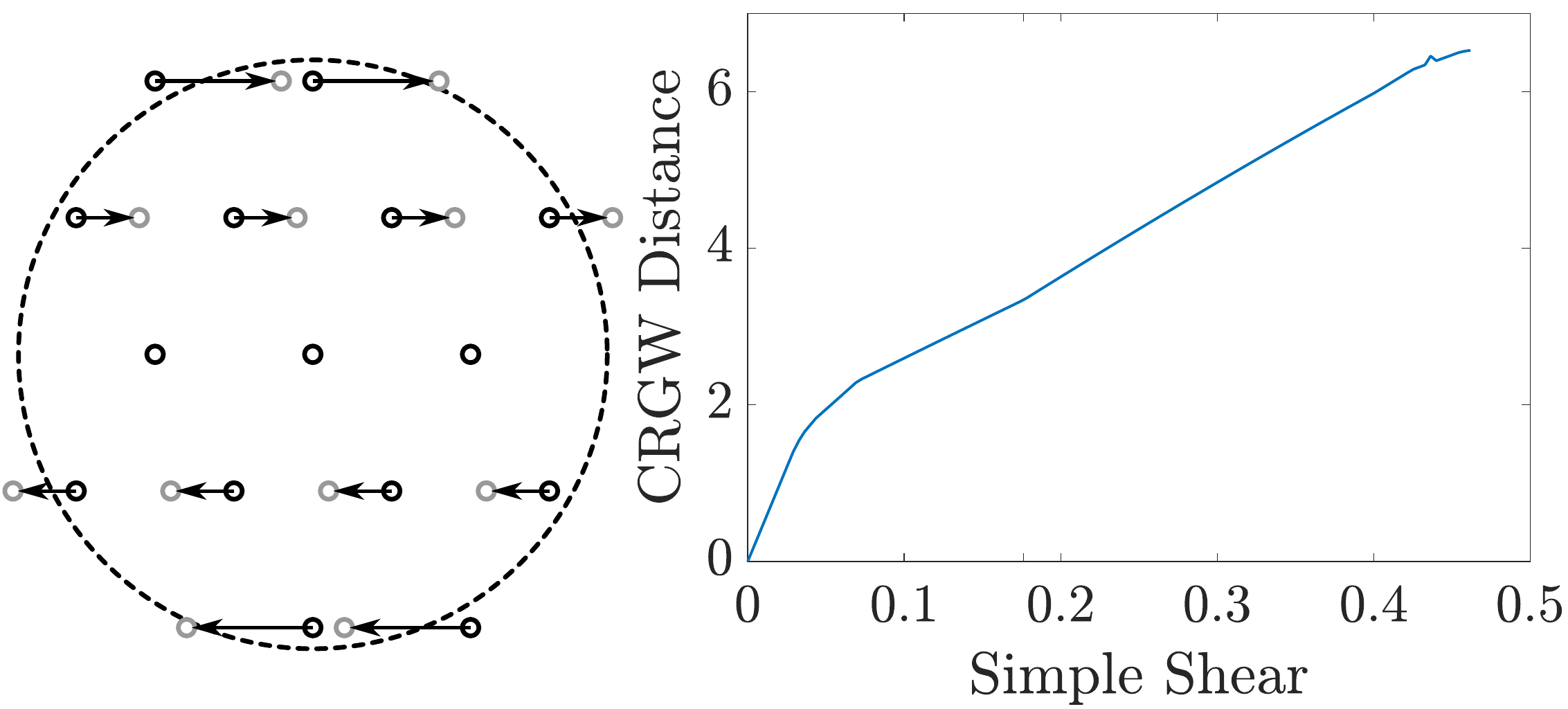}
\caption{\label{fig:change_shear}The CRGW distance is continuous with respect to elastic deformations of the local atomic environment. The environment on the left is subjected to a simple shear of $4 / (5 \sqrt{3})$. The distance to the initial condition on the right is continuous.}
\end{figure}

\begin{figure*}
\includegraphics[width=\textwidth]{%
	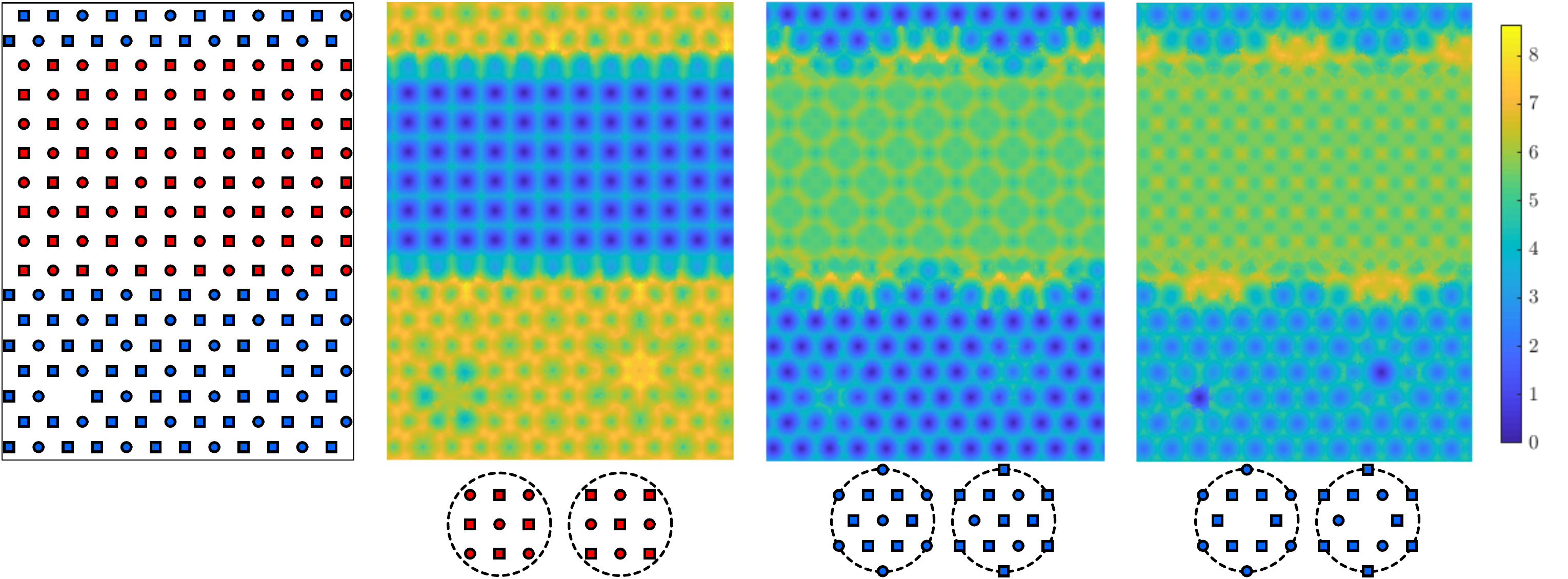}
\caption{\label{fig:multicomponent}Performance of the CRGW distance in a defected material with multiple chemical species and phases. The leftmost image shows atomic positions, with chemical species indicated by circles or squares and phases indicated by red or blue. The remaining three images show the smaller of the CRGW distances to the the two local atomic environments below the respective figure, and indicate, from left to right, atomic sites in the red phase, atomic sites in the blue phase, and vacancies in the blue phase.}
\end{figure*}

Figure \ref{fig:change_shear} shows that the CRGW distance is continuous with respect to elastic deformations of the local atomic environment. The atoms are again arranged on a triangular lattice with unit spacing, and a local atomic environment of radius $(2 + \sqrt{3}) / 2$ is subjected to a simple shear that increases to a maximum of $4 / (5 \sqrt{3})$. The CRGW distance to the initial environment is a continuous function of the shear. The shoulder around 0.5 is caused by atoms in the local environment being displaced to the boundary as the distance to the boundary decreases and the cost of matching to a reference atom increases.

The remaining figure in this section considers the performance of the CRGW distance for a defected material with multiple chemical species and phases. The leftmost image in Figure \ref{fig:multicomponent} shows the atomic positions in a simulation cell with periodic boundary conditions. The atomic shape (circle or square) and color (red or blue) indicate the chemical species and phase, where the phases can be distinguished by chemical composition and lattice type. The blue phase additionally contains two vacancies on distinct hexagonal unit cell sites. Distances are expressed in units of the interatomic spacing, which is assumed to be the same for all chemical species and phases. The radius of all local atomic environments is set to $1.75$.

The six reference atomic environments appear at the bottom of the figure, and are divided into three groups. From left to right, these correspond to atomic sites in the red phase, atomic sites in the blue phase, and vacancies in the blue phase. Local atomic environments of the same radius are constructed on a grid throughout the simulation cell, and the CRGW distances to the six reference atomic environments are calculated for each one. The right three images of Figure \ref{fig:multicomponent} show the smallest distance to any of the reference environments in the respective group, with smaller distances indicating more similarity. The atoms belonging to the red phase, the blue phase, and the interface can be identified by visual inspection of the middle images, and the location of the vacancies is clearly indicated in the rightmost image.

\section{Thermal Noise}
\label{sec:thermal_noise}

All of the examples in Section \ref{sec:classification} positioned the atoms on lattice sites, whereas molecular dynamics simulations are generally performed at finite temperatures with perturbed atomic positions. While molecular dynamics simulations can be quenched to return the atoms to their lattice sites, this requires additional computation and can complicate the observation of temperature-dependent phenomena. Hence, any approach to classify local atomic environments would ideally be robust to such perturbations. As described in Section \ref{sec:introduction}, existing geometric approaches handle this by identifying each class with some region of a feature space, with the regions defined by observation and convention rather than more fundamental considerations. This is not entirely necessary though; one could model atomic displacements as independent random variables, and derive a probability distribution of feature vectors for a given reference environment. Classification of an environment would then be reduced to, e.g., comparison with a set of prediction intervals.

This is the approach developed in the current section. Let $\XX$ be a given reference environment and $\YY$ be the same environment subject to random thermal displacements of a given magnitude. The predicted distribution of CRGW distances $\mathcal{G}(\XX, \YY)$ is constructed below, and allows one to test the hypothesis that a test environment $\ZZ$ is also derived from $\XX$ by the application of random thermal displacements. This procedure is used to classify local atomic environments in molecular dynamics simulations in Section \ref{sec:MD}.

Let there be a reference environment where all of the atoms are on the interior of the region and are not too close to the boundary. Suppose that the potential energy $\phi$ of the $i$th atom can be approximated in the vicinity of the minimum by a parabolic function
\begin{equation*}
\phi(\vec{r}_i) = \frac{1}{2} a |\vec{r}_i|^2 + b
\end{equation*}
where $\vec{r}_i$ is the atomic displacement of the $i$th atom from the position of minimum potential energy. Assuming that atomic displacements are independent, the probability distribution $p(\vec{r}_i)$ of a displacement of the $i$th atom in the canonical ensemble is a product of normal distributions
\begin{equation*}
p(\vec{r}_i) = \bigg(\frac{a}{2 \pi k_B T}\bigg)^{3/2} \exp\!\bigg({-}\frac{a |\vec{r}_i|^2}{2 k_B T}\bigg)
\end{equation*}
where $k_B$ is Boltzmann's constant and $T$ is the absolute temperature. Let $\sigma_r^2 = k_B T / a$ indicate the variance of the atomic displacements.

Let $\XX$ be a reference environment, and $\YY$ a perturbation of that environment. Suppose that the atomic perturbations are small enough that all of the $n$ atoms remain in the environment, and that each atom in $\XX$ can be unambiguously identified with an atom in $\YY$. For any natural ordering of atoms in $\XX$ and $\YY$, $\mu_{ij}$ is a diagonal matrix with ones and zeros on the diagonal. Let $\xi^{h}_{i}$ be the $i$th entry of the diagonal, with $h \in \mathcal{H}$ indicating which of the possible $2^{n}$ binary vectors is chosen. Each $\vec{\xi}^{h}$ corresponds to a particular subset of atomic pairs in $\XX$ and $\YY$ being mapped to the boundary. Equation \ref{eq:gromov_wasserstein} reduces for this case to
\begin{align}
 \mathcal{G}(\XX, \YY) &= \min_{h \in \mathcal{H}} \bigg[\frac{1}{2} \sum_{i, j}^{n} |d^X_{ij} - d^Y_{ij}| \xi_{i}^{h} \xi_{j}^{h} + \sum_{i}^{n} \lambda^{X}_{i} (1 - \xi^{h}_{i}) \nonumber \\
 &\hspace{33pt} + \sum_{i}^{n} \lambda^{Y}_{i} (1 - \xi^{h}_{i}) \bigg] \nonumber \\
 & = \min_{h \in \mathcal{H}} D_{h}.
\label{eq:covariance_gw}
\end{align}
Observe that the $D_{h}$ are correlated random variables, constructed as sums of the random variables $|d^X_{ij} - d^Y_{ij}|$ and $\lambda^{Y}_{i}$. The joint probability distribution of the $D_{h}$ will be modeled as a multivariate normal distribution using the multivariate central limit theorem. The probability distribution of $\mathcal{G}(\XX, \YY)$ can then be constructed by explicitly sampling from the joint distribution of the $D_{h}$ and finding the minimum $D_{h}$ for each sample. The problem is thereby reduced to the calculation of the means and covariance matrix of the $D_{h}$ that define the multivariate normal distribution. These are found in \ref{sec:covariance} to be
\begin{align}
 \langle D_{h} \rangle &= \sum_{i, j}^{n} \frac{\sigma_r}{\sqrt{\pi}} \xi_{i}^{h} \xi_{j}^{h} + 2 \sum_{i}^{n} \lambda^{X}_{i} (1 - \xi^{h}_{i}) \ 
\label{eq:Dmean} \\
\text{cov}(D_{h}, D_{g}) &= \bigg[ \sum_{i}^{n} (1 - \xi_{i}^{h}) (1 - \xi_{i}^{g}) \nonumber \\
 &\hspace{15pt} + \bigg(1 - \frac{2}{\pi} \bigg) \sum_{i}^{n}  \sum_{j \ne i}^{n} \xi_{i}^{h}  \xi_{j}^{h} \xi_{i}^{g} \xi_{j}^{g} \nonumber\\
 &\hspace{15pt} + \sum_{i}^{n}  \sum_{j \ne i}^{n} \sum_{k \ne i, j}^{n}  \xi_{i}^{h} \xi_{j}^{h} \xi_{i}^{g} \xi_{k}^{g} f(\theta_{ijk}) \bigg] \sigma_r^2
 \label{eq:Dcovariance}
\end{align}
where $\langle \cdot \rangle$ indicates the mean of a quantity and $f{(\theta_{ijk})}$ is defined by  Eq.\ \ref{eq:cov_int_int}. The covariance matrix depends on the geometry of the reference environment via the angles $\theta_{ijk}$ between triplets of atoms in the reference environment.

\begin{figure}
\centering
\includegraphics[width=\columnwidth]{%
	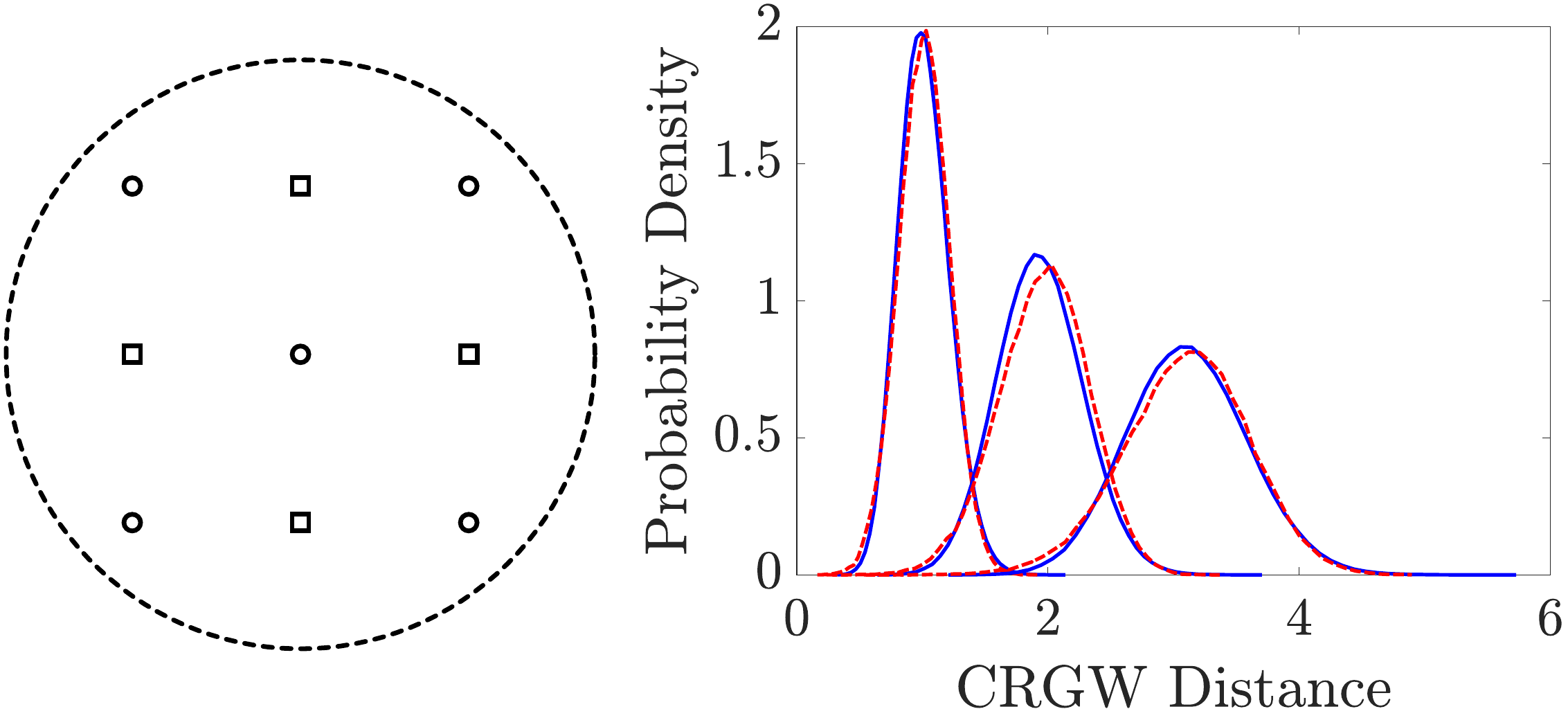}
\caption{\label{fig:multiple_temperatures}The local atomic environment on the left contains five circle atoms and four square atoms on a square lattice with unit spacing in a region of radius $(\sqrt{2}+ 2) / 2$. The measured (solid blue) and predicted (dashed red) distributions of CRGW distances for $\sigma_r = 0.025$, $0.05$, and $0.1$ are on the right.}
\end{figure}

To sample from this distribution, define the matrix elements $\Sigma_{hg} = \text{cov}(D_h, D_g)$ and find any real matrix $\mat{A}$ such that $\mat{\Sigma} = \mat{A}\mat{A}^T$. Let $z_g$ be a random variable distributed according to the standard normal distribution. Then
\begin{equation*}
Y = \min_{h \in \mathcal{H}} \bigg[ \langle D_h \rangle + \sum_g A_{hg} z_g \bigg] \
\end{equation*}
samples from the distribution of $\mathcal{G}(\XX, \YY)$ implicitly defined by Eq.\ \ref{eq:covariance_gw}.

Figure \ref{fig:multiple_temperatures} provides some numerical evidence that samples of $\mathcal{G}(\XX, \YY)$ can be used to construct the empirical distribution. The reference environment on the left resembles one in Figure \ref{fig:multicomponent}. The plot on the right shows that the predicted probability distribution (dashed red) is a good approximation for the measured one (solid blue), even when the standard deviations of the atomic displacements are as large as one-tenth the average atomic spacing. The small offset of the mean is likely the result of three sources of error; the atomic displacements are assumed to be small relative to the atomic spacing, $D_h$ is a sum of random variables that are not identically distributed, and the number of random variables in some of the $D_h$ is relatively small.

\section{Applications to Molecular Dynamics}
\label{sec:MD}

This section describes the use of the CRGW distance to classify atomic environments in several molecular dynamics (MD) simulations performed in LAMMPS \cite{1993plimpton}. The initial application compares the ability of the CRGW distance to distinguish simple crystal structures (i.e., BCC, FCC, and HCP) with that of ACNA and PTM. To that end, BCC tungsten \cite{2013marinica}, FCC copper \cite{2015pascuet}, and HCP magnesium \cite{2015wu} single crystals were simulated at temperatures up to melting in the isothermal-isobaric ensemble (NPT). The simulated systems respectively contained $4394$, $8788$, and $8788$ atoms. The BCC and FCC unit cells were cubic, while the non-standard HCP unit cell was length $a$ in the $x$-direction, $\sqrt{3} a$ in the $y$-direction, and $\sqrt{8/3} a$ in the $z$-direction. A single crystal of each material was quenched to $0 \unit{K}$, then heated in increments of $20 \unit{K}$ up to melting with an equilibration of $3 \unit{ps}$ at each temperature. The exceptions to this are that tungsten was heated in increments of $50 \unit{K}$ and equilibrated for $5 \unit{ps}$ ps above $4000 \unit{K}$, and aluminum and magnesium were heated in increments of $10 \unit{K}$ below $300 \unit{K}$. The pressure was set to $0 \unit{bar}$ throughout.

\begin{figure}%
	\centering
	\includegraphics[width=\columnwidth]{%
	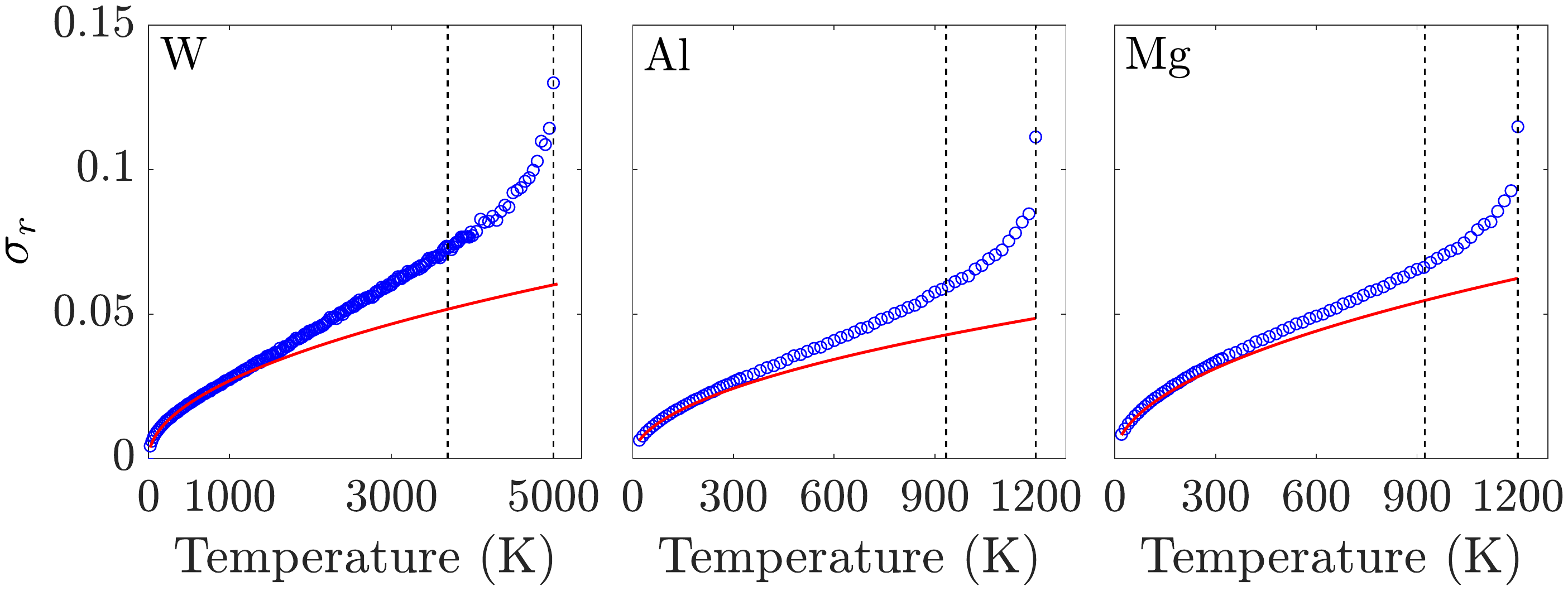}
\caption{\label{fig:sigma_r}Blue circles show $\sigma_r$ measured from simulations, and red curves show a $\sigma_r \propto \sqrt{T}$ trend line fit. For each figure, the dashed vertical line to the left indicates the true melting point of the crystal, and the dashed vertical line to the right indicates the apparent melting point for an average heating rate of $6.66 \times 10^{12} \unit{K / s}$.}
\end{figure}

The $\sigma_r$ values used to construct the predicted CRGW distance distributions were found by directly measuring atomic displacements in the MD simulations after accounting for translation, rotation, and expansion of the local environments. This is effectively a measure of the magnitude of thermal displacements, and is predicted in Section \ref{sec:thermal_noise} to increase as $\sqrt{T}$. Figure \ref{fig:sigma_r} shows that $\sigma_r$ follows this expectation reasonably accurately for temperatures below one-third of the melting point. Lindemann's criterion \cite{2007chakravarty, 2017sarkar} further suggests that melting occurs if $\sigma_r$ exceeds a critical value. The melting points of the potentials were identified by discontinuities in the potential energy per atom at $5000$ K for tungsten, $1200$ K for aluminum, and $1200$ K for magnesium, and generally occurred when $\sigma_r \approx 0.1$ in units of the average atomic spacing.

The classification of atomic environments in this section is based on $p$-values, or the probability of obtaining a CRGW distance at least as extreme as the one observed given that the local atomic environment actually derives from the specified reference environment. If the CRGW distance falls below the median of a predicted distribution, the mass of the predicted distribution to the left of that distance is the $p$-value. If the CRGW distance falls above the median reference environment, the mass of the predicted distribution to the right of that distance is the $p$-value. The maximum possible $p$-value is $0.5$ when the CRGW distance is exactly the median value. 

\begin{figure}
\centering
\includegraphics[width=\columnwidth]{%
	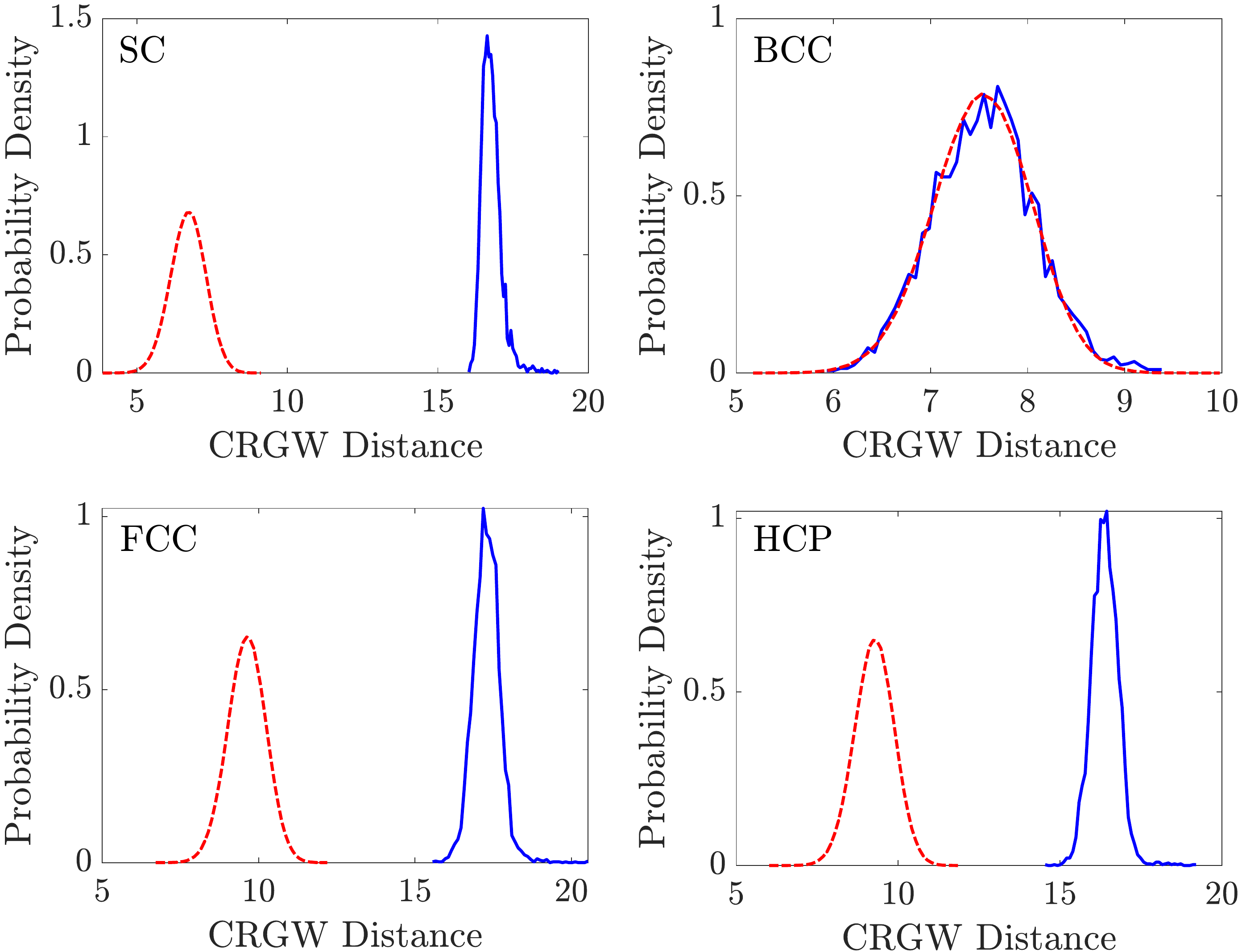}
\caption{\label{fig:tungsten1000K}The CRGW distances calculated for tungsten local environments at $1000 \unit{K}$ (solid blue lines) compared to the distributions predicted for SC, BCC, FCC, and HCP reference environments (dashed red lines). $97.2\%$ of tungsten atoms were correctly classified as BCC at this temperature.}
\end{figure}

Consider the classification of atomic environments in tungsten at $1000 \unit{K}$ in Figure \ref{fig:tungsten1000K}. The $p$-values for each local environment were calculated for SC, BCC, FCC, and HCP reference environments. If a local environment's $p$-value was greater than $0.01$ for a particular reference environment and was lower for all other reference environments, then the local environment was classified accordingly. This two-part condition ensures that, e.g., the environment sufficiently resembles a perturbed BCC environment and is more likely to be a perturbed BCC environment than any other reference environment. This classification scheme is much more rigorous than those used in the past, and effectively provides the user with an uncertainty in addition to the classification. 

Figure \ref{fig:tungsten1000K} more specifically plots the measured CRGW distance distributions between a local atomic environment and a given reference environment (solid blue), and the probability distributions that would be predicted if the local atomic environment really were a perturbation of that reference environment (dashed red). That the probability distributions coincide for the BCC structure indicates that the vast majority of atoms should be classified as BCC. For this particular simulation, $97.2 \%$ of atoms were correctly classified as BCC.

\begin{figure}
\centering
	\includegraphics[width=\columnwidth]{%
	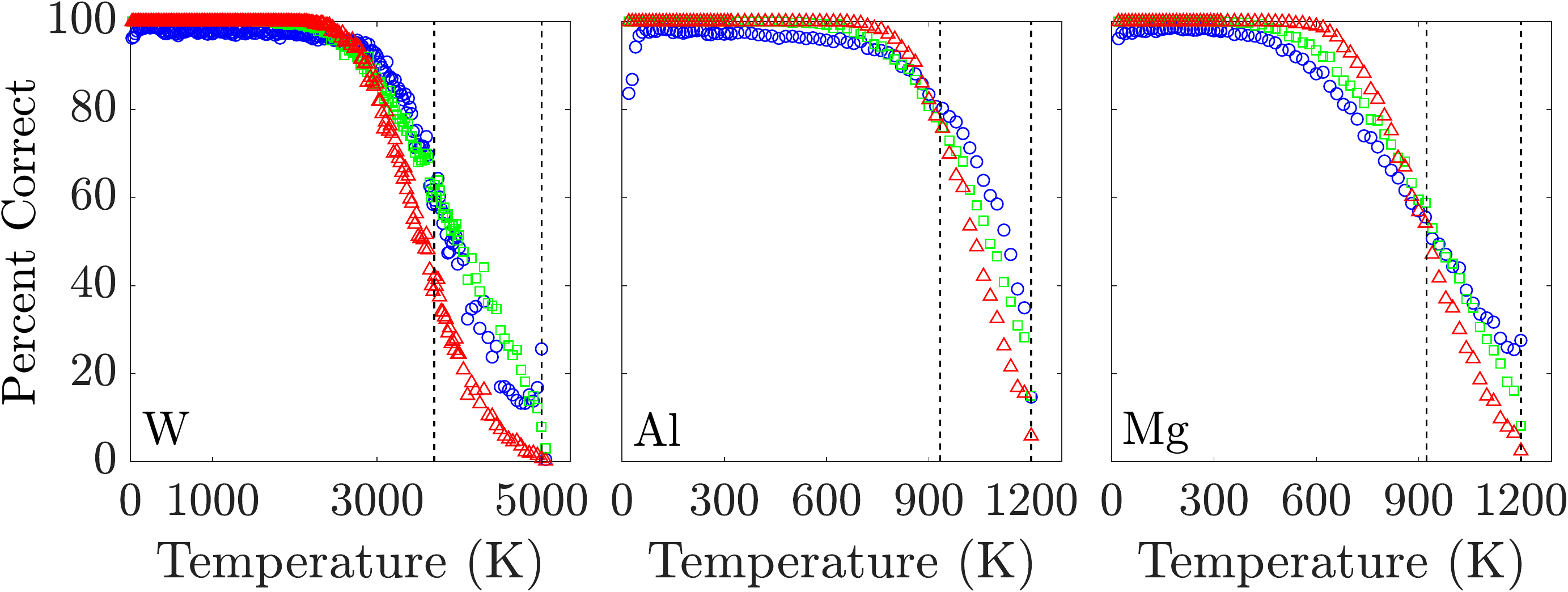}
\caption{\label{fig:percent}Percent of tungsten classified as BCC (left), percent of aluminum classified as FCC (middle), and percent of magnesium classified as HCP (right) as a function of temperature. Blue circles are for the CRGW distance with a $p$-value criterion of $0.01$, green squares are for ACNA, and red triangles are for PTM. For each figure, the dashed vertical line to the left indicates the true melting point of the crystal, and the dashed vertical line to the right indicates the melting point of the potential.}
\end{figure}

Figure \ref{fig:percent} shows the percent of tungsten classified as BCC, the percent of aluminum classified as FCC, and the percent of magnesium classified as HCP as functions of temperature. The classification scheme described in this section (blue circles) correctly classifies more than $95\%$ of the atoms up to two-thirds of the melting point for BCC and FCC. The slight dip at the lower temperatures are perhaps due to low-frequency phonons being mistaken as rotations in the measurement of $\sigma_r$, and the earlier decline for the HCP structure could be caused by the approximation of spherically-symmetric atomic displacements being less valid for noncentrosymmetric materials. Nevertheless, the method correctly classifies more than $90\%$ of the atoms at half the melting point for HCP. This performance is comparable to that of ACNA (green squares) and PTM (red triangles) as implemented in OVITO \cite{ovito}, though the CRGW distance is considerably more expensive to calculate. Specifically, informal measurements suggest that ACNA, PTM, and the CRGW distance require $1 \unit{\mu s / atom}$, $5 \unit{\mu s / atom}$, and $0.25 \unit{s / atom}$ to classify atomic environments. That is, the CRGW distance is roughly $10^5$ times slower, making real-time analysis impractical.

\begin{figure*}
\centering
	\includegraphics[width=\textwidth]{%
	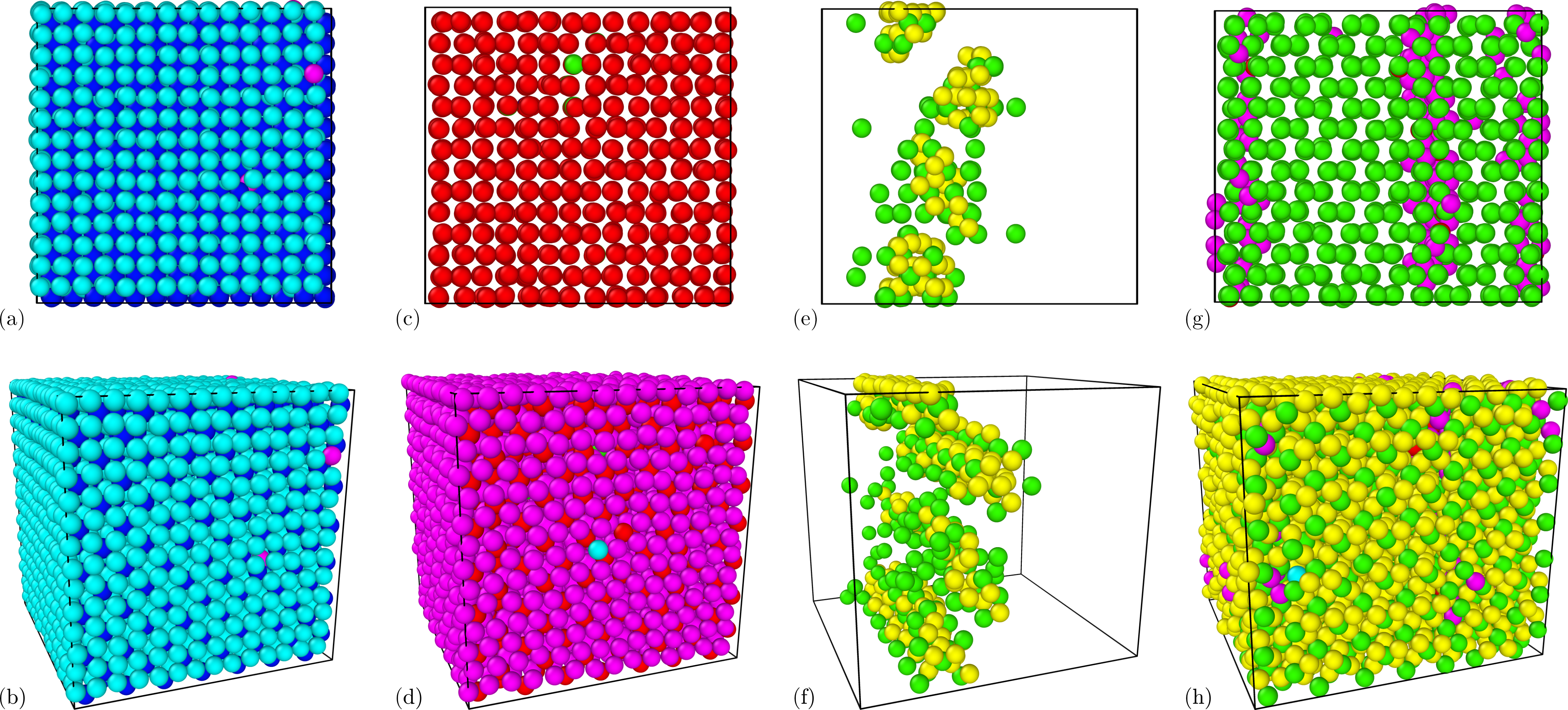}
\caption{\label{fig:zirconia}Molecular dynamics simulation of a phase transformation from cubic to monoclinic zirconia. Zirconium and oxygen atoms are dark blue and light blue in the cubic phase, green and yellow in the monoclinic phase, and red and purple otherwise. The $[100]$ and $[010]$ directions of the cubic and monoclinic phases are respectively to the left and out of the page in the top row. (a-b) Cubic zirconia at $786 \unit{K}$. (c-d) Intermediate structure at $812 \unit{K}$. Only zirconium is shown in (c) to reveal the incipient symmetry breaking. (e-f) The monoclinic phase is nucleated at $812 \unit{K}$, $1.1 \unit{ps}$ after the structure in (c-d). (g-h) The transformation is completed at $1400 \unit{K}$, $62.5 \unit{ps}$ after the structure in (e-f). Only zirconium in the monoclinic phase and unclassified oxygen is shown in (g) to reveal the interfacial defects.}
\end{figure*}

The utility of the CRGW distance instead lies in the ability to classify more complicated atomic environments. Yu et al.\ proposed an interatomic potential for zirconia \cite{2009yu} modeled on the well-known BKS potential for silica \cite{1990vanbeest}. They found the potential to be suitable for simulations of cubic and monoclinic zirconia, with the monoclinic phase being slightly preferred by $0.11 \unit{eV / ZrO_2}$ (the tetragonal phase spontaneously transforms to cubic). A simulation cell containing a single crystal of cubic zirconia with seven unit cells along each coordinate direction was prepared and relaxed at $0 \unit{K}$ and $0 \unit{bar}$. The simulation then proceeded in the isothermal-isobaric (NPT) ensemble, with the temperature raised in intervals of $28 \unit{K}$ every $3 \unit{ps}$ and the pressure maintained at $0 \unit{bar}$. Given the lower enthalpy of the monoclinic phase, a phase transformation from cubic to monoclinic was expected. CRGW distances were calculated for zirconium- and oxygen-centered reference environments of radius $4.04 \,\text{\AA}$ in the cubic and monoclinic phases relaxed at $0 \unit{K}$ and $0 \unit{bar}$. While the approximation that the zirconium and oxygen atoms experience the same magnitude thermal vibrations is poor, the same classification criterion was used as above with $p$-values in the interval of $10^{-2}$ to $10^{-4}$ depending on the phase of interest.

The expected transformation occurred in three stages, shown in Figure \ref{fig:zirconia}. The cubic phase in (a-b) remained stable up to $786 \unit{K}$, with distributed disorder developing over a period of $2.9 \unit{ps}$ to give (c-d). This involved $[010]$ columns of zirconium atoms displacing along $[100]$ directions, as revealed by (c) where only zirconium atoms are shown. The disordered structure subsequently developed three monoclinic nuclei over a period of $1.1 \unit{ps}$, visibly extending along the $[010]$ direction in (e-f). The positioning of the nuclei suggests that they are not energetically independent, but interact mechanically as a consequence of the transformation strain and the periodic boundary conditions. These remained stable for several tens of picoseconds, but eventually merged and grew to give the monoclinic system in (g-h) after $62.5 \unit{ps}$. The transformation did not result in a single crystal though, with two distinct regions differing by a non-lattice translation in the $[010]$ direction. These regions can be identified in (g) either by the pattern of the columns of zirconium atoms or by unclassified oxygen atoms that occur at the interfaces. The existence of these interfaces is intimately related to the use of periodic boundary conditions, and should not be construed as a general feature of the transition.

Indeed, a careful study of the cubic to monoclinic zirconia phase transition would require investigating size effects, homogeneous and heterogeneous nucleation barriers, the elastic strains in the transformed structure, and the slight differences between the relaxed monoclinic structure and that published in the literature \cite{2006whittle}. This is not undertaken here since the purpose of this study is instead to show that the CRGW distance can be used to classify atomic environments in systems at elevated temperatures with more species and more complicated crystal structures than can be handled by standard ACNA and PTM.

\section{Conclusion}
\label{sec:conclusion}

An automated method to classify local atomic environments via the composition-restricted Gromov--Wasserstein (CRGW) distance is proposed. Advantageous properties of this method include that it is invariant to translations, rotations, and reflections of the local atomic environment, and that it does not require the local atomic environment to be centered on an atom. The method does not make any assumption about the material class, making it applicable with minimal modification to materials with multiple chemical species and general crystal structures. Molecular dynamics results for single crystals verify that the method is a reliable approach to classifying local atomic environments in pure metals at temperatures up to half the melting point, albeit less efficiently than for techniques already available in the literature. The strength of the method is instead its applicability to general atomic systems, as is demonstrated by preliminary analysis of a cubic to monoclinic phase transition in zirconia.

\section*{Acknowledgements}

J.K.M.\ was supported by the National Science Foundation under Grant No.\ DMR 2003849.

\appendix

\section{Heuristics}
\label{sec:heuristics}

The minimization problem in Eq.\ \ref{eq:gromov_wasserstein} is difficult because of the presence of many local minima, some of them introduced by symmetries in the reference environment. Specifically, the algorithm described in Section \ref{sec:gromov_wasserstein} can split an atom's mass between several reference atoms related by a symmetry operation. The implementation handles this by forcefully breaking the symmetry and assigning the first such atom to precisely one other atom after each step of alternate convex search. This gradually forces the coupling matrix to be a (0, 1)-matrix, where the atoms of the reference and local structures are either matched or sent to the boundary and partial matching is disallowed. Second, the algorithm for the unbalanced GW distance often finds a local minimum by sending all atoms to the boundary. This is discouraged by beginning with artificially high values of $\vec{\lambda}^X$ and $\vec{\lambda}^Y$ in Eq.\ \ref{eq:gromov_wasserstein}, and gradually relaxing them to their final values. Third, a central atom is sometimes inserted with a species label that differs from all other atoms in the environment. Forcing the center atom in the reference environment to be assigned to that in the local environment empirically helps the other atoms to be assigned consistently. While the resulting algorithm cannot guarantee a unique minimum distance coupling, the results in Sections \ref{sec:thermal_noise} and \ref{sec:MD} strongly suggest that the minimum is achieved in almost every case.

\section{Mean and Covariance of the $D_{h}$}
\label{sec:covariance}

Initially consider $\langle D_{h} \rangle$, the mean of $D_{h}$ for normally-distributed atomic displacements. From the definition in Eq.\ \ref{eq:covariance_gw}, the relevant equation is
\begin{align*}
 \langle D_{h} \rangle &= \frac{1}{2} \sum_{i, j}^{n} \langle |d^X_{ij} - d^Y_{ij}| \rangle \xi_{i}^{h} \xi_{j}^{h} + \sum_{i}^{n} \lambda^{X}_{i} (1 - \xi^{h}_{i})  \nonumber\\
 & \ \quad + \sum_{i}^{n} \langle \lambda^{Y}_{i} \rangle (1 - \xi^{h}_{i}) \
\end{align*}
where $\lambda^{X}_i$ is the constant distance to the boundary in the reference environment. As described in Section \ref{sec:thermal_noise}, the probability distribution $p(\vec{r}_i)$ of a displacement of the $i$th atom in the canonical ensemble is assumed to be
\begin{equation*}
p(\vec{r}_i) = \bigg(\frac{1}{2 \pi \sigma_r^2}\bigg)^{3/2} \exp\!\bigg({-}\frac{|\vec{r}_i|^2}{2 \sigma_r^2}\bigg)
\end{equation*}
with $\sigma_r^2 = k_B T / a$ indicating the variance of the atomic displacements.

Since $p(\vec{r}_i)$ is spherically symmetric and $|\lambda^{Y}_i - \lambda^{X}_i| \ll \lambda^{X}_i$ is assumed, the distance to the external boundary is distributed as 
\begin{equation*}
p(\lambda^{Y}_i) = \bigg(\frac{1}{2 \pi \sigma_r^2}\bigg)^{1/2} \exp\!\bigg({-}\frac{ (\lambda^{Y}_i - \lambda^{X}_i)^2}{2 \sigma_r^2}\bigg),
\end{equation*}
from which the mean and variance of $\lambda^{Y}_i$ are found to be
\begin{align}
\langle \lambda^{Y}_i \rangle &= \lambda^{X}_i
\label{eq:lambdamean}  \\
\text{var}(\lambda^{Y}_i)& = \sigma_r^2.
\label{eq:lambdavar}
\end{align}
This specifies the terms in the third sum in the equation for $\langle D_{h} \rangle$ above.

\begin{figure}
\centering
\includegraphics[width=0.7\columnwidth]{%
	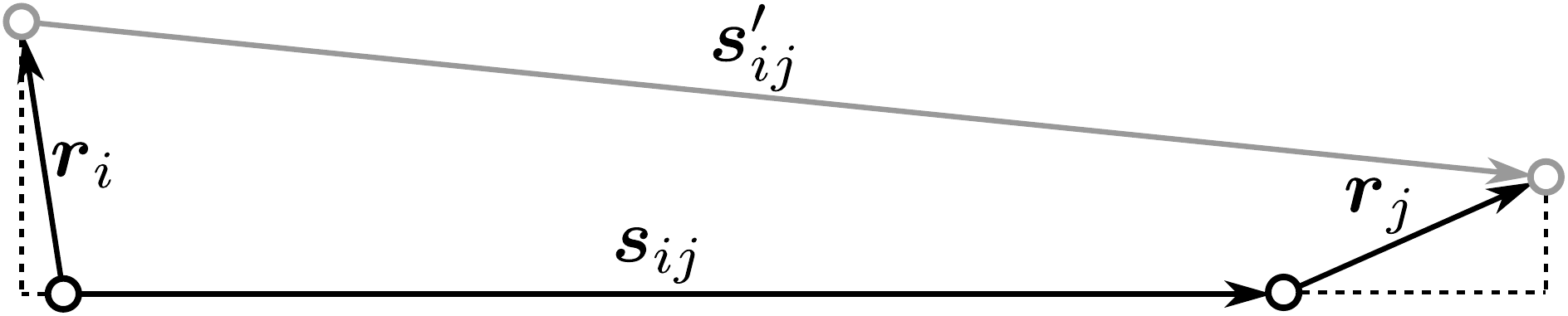}
\caption{\label{fig:displacements}If the lengths of $\vec{r}_i$ and $\vec{r}_j$ are small relative to $\vec{s}_{ij}$, then $\vec{s}_{ij}$ and $\vec{s}'_{ij}$ are nearly parallel, and the difference in the length of $\vec{s}'_{ij}$ and $\vec{s}_{ij}$ is approximately the difference of the projections of $\vec{r}_j$ and $\vec{r}_i$ onto $\vec{s}_{ij}$.}
\end{figure}

Now consider $|\delta_{ij}| = |d^X_{ij} - d^Y_{ij}|$ for  $\textit{i} \ne \textit{j}$. As described in Section \ref{sec:gromov_wasserstein}, the GW distance between the reference environment and a perturbed environment is effectively the sum of the magnitudes of the changes in the distances between all pairs of atoms. Let $\vec{s}_{ij}$ be the vector from the $i$th atom to the $j$th atom; the central quantity of interest is the change in the length of this vector with the application of perturbations. Figure \ref{fig:displacements} suggests that if the perturbations are small relative to $\vec{s}_{ij}$, then
\begin{equation*}
\delta_{ij} = (\vec{r}_j - \vec{r}_i) \cdot \nvec{s}_{ij}
\end{equation*}
is the approximate change in the length of $\vec{s}_{ij}$, where $\nvec{s}_{ij}$ is the unit vector $\vec{s}_{ij} / |\vec{s}_{ij}|$. Since $p(\vec{r}_i)$ is spherically symmetric, the probability distribution of the projected displacement $\vec{r}_i \cdot \nvec{s}_{ij}$ is the normal distribution
\begin{equation}
p(\vec{r}_i \cdot \nvec{s}_{ij}) = \frac{1}{\sqrt{2 \pi \sigma_r^2}} \exp\!\bigg({-}\frac{|\vec{r}_i \cdot \nvec{s}_{ij}|^2}{2 \sigma_r^2}\bigg).
\label{eq:normal}
\end{equation}
The probability distribution $p(\delta_{ij})$ can be found from Eq.\ \ref{eq:normal} by a change of variables; if $\epsilon_{ij} = (\vec{r}_j + \vec{r}_i) \cdot \nvec{s}_{ij}$ is the counterpart to $\delta_{ij}$, and $p(\vec{r}_j \cdot \nvec{s}_{ij}) p(\vec{r}_i \cdot \nvec{s}_{ij})$ is the joint distribution of $\vec{r}_j \cdot \nvec{s}_{ij}$ and $\vec{r}_i \cdot \nvec{s}_{ij}$, then
\begin{align*}
p(\delta_{ij}, \epsilon_{ij}) &= \frac{1}{2} p\bigg(\frac{\epsilon_{ij} + \delta_{ij}}{2}\bigg) p\bigg(\frac{\epsilon_{ij} - \delta_{ij}}{2}\bigg) \\
&= \frac{1}{4 \pi \sigma_r^2} \exp\!\bigg({-}\frac{\epsilon_{ij}^2 + \delta_{ij}^2}{4 \sigma_r^2}\bigg)
\end{align*}
is the joint distribution of $\epsilon_{ij}$ and $\delta_{ij}$, where the factor of $1/2$ is the Jacobian determinant of the transformation. Integrating over $\epsilon_{ij}$ and observing that $p(\delta_{ij})$ is a symmetric function gives
\begin{equation*}
p(|\delta_{ij}|) = \frac{1}{\sqrt{\pi \sigma_r^2}} \exp\!\bigg({-}\frac{|\delta_{ij}|^2}{4 \sigma_r^2}\bigg)
\end{equation*}
for the probability distribution of the magnitude of the change in the distance between the $i$th and $j$th atoms. The resulting mean and variance are
\begin{align}
\langle |\delta_{ij}| \rangle &= \frac{2 \sigma_r}{\sqrt{\pi}}
\label{eq:deltamean} \\
\text{var}(|\delta_{ij}|) &= \bigg( 2 - \frac{4}{\pi} \bigg) \sigma_r^2. 
\label{eq:deltavar}
\end{align}
Using Eqs.\ \ref{eq:lambdamean} and \ref{eq:deltamean} allows the equation for $\langle D_{h} \rangle$ introduced at the beginning of this section to be reduced to Eq.\ \ref{eq:Dmean}.

This only leaves the calculation of the covariance matrix. From the definition of the covariance:
\begin{equation*}
\text{cov}(D_{h}, D_{g}) =  \langle D_{h}D_{g} \rangle -  \langle D_{h}\rangle \langle D_{g} \rangle.
\end{equation*}
Expanding all the products and cancelling terms gives
\begin{align*}
\text{cov}(D_{h}, D_{g}) & = \frac{1}{4} \sum_{i, j}^{n} \sum_{i', j'}^{n} \xi_{i}^{h} \xi_{j}^{h} \xi_{i'}^{g} \xi_{j'}^{g} \text{cov}(|\delta_{ij}| , |\delta_{i'j'}|) \nonumber\\
 & \ \quad + \frac{1}{2} \sum_{i, j}^{n} \sum_{i'}^{n} \xi_{i}^{h} \xi_{j}^{h} (1 - \xi_{i'}^{g}) \text{cov}(|\delta_{ij}| , \lambda^{Y}_{i'}) \nonumber\\
 & \ \quad + \frac{1}{2} \sum_{i', j'}^{n} \sum_{i}^{n} \xi_{i}^{g} \xi_{j}^{g} (1 - \xi_{i'}^{h}) \text{cov}(\lambda^{Y}_{i}, |\delta_{i'j'}|) \nonumber\\
 & \ \quad + \frac{1}{2} \sum_{i, i'}^{n} (1 - \xi_{i}^{h}) (1 - \xi_{i'}^{g}) \text{cov}(\lambda^{Y}_{i} , \lambda^{Y}_{i'}).
\end{align*}
We start with the last term. $\lambda^{Y}_{i}$ and $\lambda^{Y}_{i'}$ for $i \ne i'$ are independent by inspection, so this reduces to $\sum_{i} (1 - \xi^{h}_{i})(1 - \xi^{g}_{i}) \sigma_r^2$ by Eq.\ \ref{eq:lambdavar}. Now consider $\text{cov}(|\delta_{ij}| , |\delta_{i'j'}|)$. If $i = i'$  and $j = j'$, then this reduces to $\text{var}(|\delta_{ij}|)$ as given in Eq.\ \ref{eq:deltavar}. If all the indices are distinct, then $\text{cov}(|\delta_{ij}| , |\delta_{i'j'}|)$ vanishes by inspection. The only remaining case is for $\text{cov}(|\delta_{ij}| , |\delta_{ik}|)$ for $\textit{j} \ne \textit{k}$.

\begin{figure}
\centering
\includegraphics[width=0.55\columnwidth]{%
	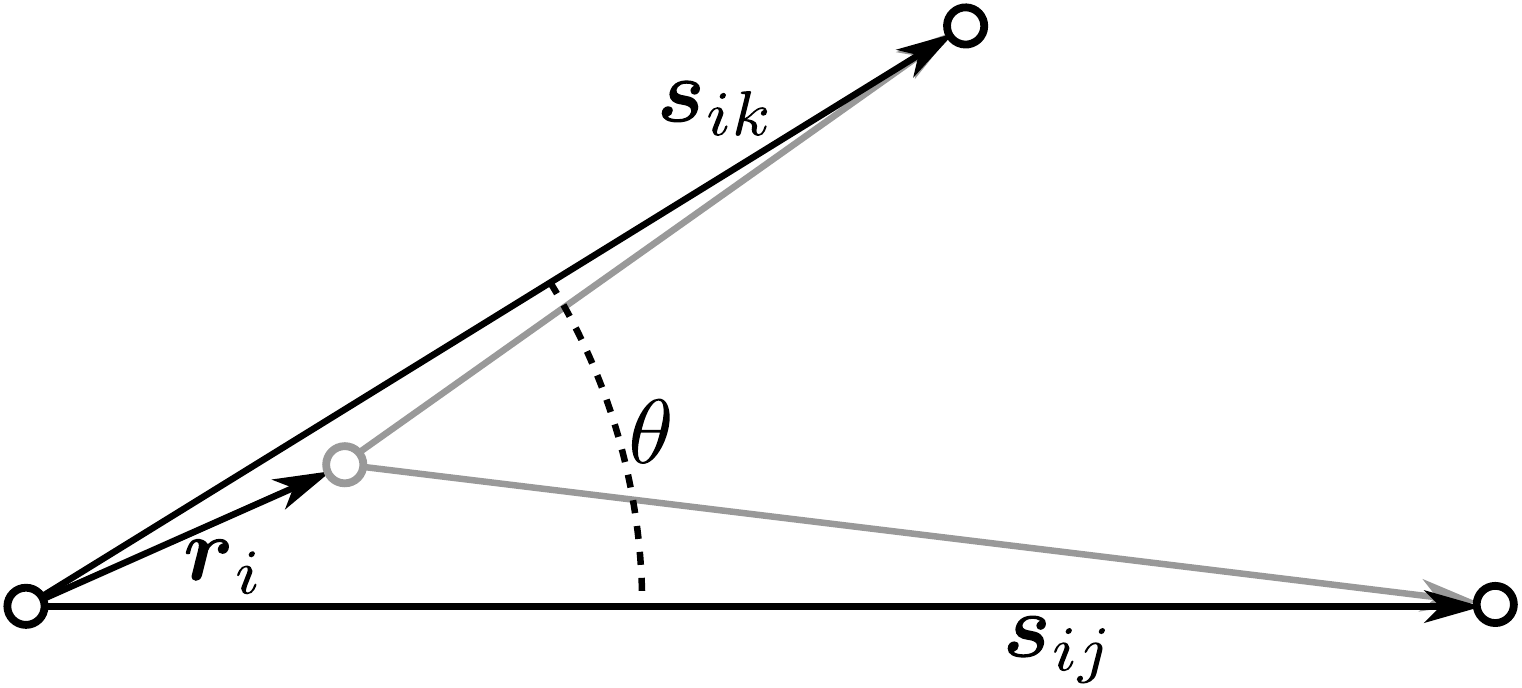}
\caption{\label{fig:correlation}The changes in the lengths of the vectors $\vec{s}_{ij}$ and $\vec{s}_{ik}$ with a displacement $\vec{r}_i$ are correlated, with the strength of the correlation depending on the angle $\theta$.}
\end{figure}

With reference to Figure \ref{fig:correlation}, a coordinate system is constructed in the plane of the page with the $x$-axis along $\nvec{s}_{ij}$ and the $y$-axis in the vertical direction. The joint distribution of the  $\zeta = \vec{r}_i \cdot \nvec{s}_{ij}$ and $\eta = \vec{r}_i \cdot \nvec{s}_{ik}$ is found from that of $x$ and $y$ by the change of variables
\begin{equation*}
x= \zeta \quad\quad y = -\cot(\theta)\zeta + \csc(\theta)\eta
\end{equation*}
with the Jacobian determinant $\csc(\theta)$. The resulting distribution is
\begin{equation*}
p(\zeta, \eta) = \frac{\csc{\theta}}{2 \pi \sigma_r^2} \exp\!\bigg[{-}\frac{\csc{\theta}^2 (\zeta^2 - 2 \cos{\theta \zeta \eta} + \eta^2)}{2 \sigma_r^2} \bigg].
\end{equation*}
This is multiplied by a normal distribution of $\vec{r}_j \cdot \nvec{s}_{ij}$ like the one in Eq.\ \ref{eq:normal}, a change of variables
\begin{equation*}
\vec{r}_i \cdot \nvec{s}_{ij} = (\epsilon_{ij} - \delta_{ij}) / 2 \quad\quad \vec{r}_j \cdot \nvec{s}_{ij} = (\epsilon_{ij} + \delta_{ij}) / 2 
\end{equation*}
with the Jacobian determinant $1 / 2$ is performed, and the dependence on $\epsilon_{ij}$ is integrated out to find the joint distribution of $\delta_{ij}$ and $\vec{r}_i \cdot \nvec{s}_{ik}$. This procedure is repeated with $\vec{r}_k \cdot \nvec{s}_{ik}$ to find the joint distribution of $\delta_{ij}$ and $\delta_{ik}$:
\begin{align*}
p(\delta_{ij}, \delta_{ik}) &= \frac{1}{\pi \sqrt{14 - 2 \cos(2 \theta)} \sigma_r^2} \\
&\quad\; \exp\!\bigg\{{-}\frac{2 [\delta_{ij}^2 + \delta_{ik}^2 - \delta_{ij} \delta_{ik} \cos(\theta)]}{[7 - \cos(2 \theta)] \sigma_r^2}\bigg\}.
\end{align*}
The joint distribution of $|\delta_{ij}|$ and $|\delta_{ik}|$ is constructed from $p(\delta_{ij}, \delta_{ik})$ by adding together the four variants with each combination of signs for $\delta_{ij}$ and $\delta_{ik}$. Given $p(|\delta_{ij}|, |\delta_{ik}|)$, the covariance of $|\delta_{ij}|$ and $|\delta_{ik}|$ is found to be
\begin{align}
\text{cov}(|\delta_{ij}|, |\delta_{ik}|) &= \frac{1}{\pi} \bigg\{2 \arctan\bigg[\frac{\sqrt{2} \cos(\theta)}{\sqrt{7 - \cos(2 \theta)}}\bigg] \cos(\theta) \nonumber \\
&\phantom{= \frac{1}{\pi} \bigg\{ } + \sqrt{14 - 2 \cos(2 \theta)}- 4\bigg\} \sigma_r^2 \nonumber \\
&= f(\theta) \sigma_r^2. 
\label{eq:cov_int_int}
\end{align}

The remaining terms in the equation for $\text{cov}(D_{h}, D_{g})$ are those involving $\text{cov}(|\delta_{ij}|,\lambda^{Y}_{i'})$. Since this vanishes by inspection for $i' \ne \{\textit{i}, \textit{j}\}$, only $\text{cov}(|\delta_{ij}|,\lambda^{Y}_{i})$ need be considered further. Suppose that the probability of the $i$th atom leaving the environment is vanishing small. Then a procedure analogous to that followed for $p(\delta_{ij}, \delta_{ik})$ gives 
\begin{align*}
p(\delta_{ij}, \lambda^{Y}_{i}) &= \frac{1}{\pi \sqrt{6 - 2 \cos(2 \theta)} \sigma_r^2} \\
&\quad\; \exp\!\bigg\{{-}\frac{\delta_{ij}^2 + 2 \omega_{i}^2 - 2 \delta_{ij} \omega_{i} \cos(\theta)}{[\cos(2 \theta) - 3] \sigma_r^2}\bigg\}
\end{align*}
for the joint distribution of $\delta_{ij}$ and $\lambda^{Y}_{i}$, where $\omega_{i} = \lambda^{Y}_{i} - \lambda^{X}_{i}$. The joint distribution of $|\delta_{ij}|$ and $\lambda^{Y}_{i}$ is constructed from $p(\delta_{ij}, \delta_{i})$ by adding the two variants with each sign of $\delta_{ij}$. Remarkably, the covariance of $|\delta_{ij}|$ and $\lambda^{Y}_{i}$ is found to vanish. 

At this point, the covariance of $D_h$ and $D_g$ can be given explicitly as 
\begin{align*}
\text{cov}(D_{h}, D_{g}) & =  \sum_{i}^{n} (1 - \xi_{i}^{h}) (1 - \xi_{i}^{g}) \sigma_r^2 \nonumber\\
 & \ \quad + \frac{1}{2} \sum_{i', j'}^{n}  \xi_{i}^{h}  \xi_{j}^{h} \xi_{i}^{g} \xi_{j}^{g} \text{var}(|\delta_{ij}|) \nonumber\\
 & \ \quad + \sum_{i, j}^{n} \sum_{k \ne j}^{n}  \xi_{i}^{h} \xi_{j}^{h}  \xi_{i}^{g} \xi_{k}^{g} \text{cov}(|\delta_{ij}|, |\delta_{ik}|).
\end{align*}
where the multipliers for the second and third terms arise from the number of ways to assign the shared indices. Substituting Eqs.\ \ref{eq:deltavar} and \ref{eq:cov_int_int} for $\text{var}(|\delta_{ij}|)$ and $\text{cov}(|\delta_{ij}|, |\delta_{ik}|)$ then gives Eq.\ \ref{eq:Dcovariance}.

\section*{Data availability}
\label{sec:data_availability}

The raw data required to reproduce these findings cannot be shared at this time due to technical or time limitations. The processed data required to reproduce these findings cannot be shared at this time due to technical or time limitations.

\bibliography{refs}

\end{document}